\documentclass[reviewcopy]{elsart}

\usepackage{adnd}

\usepackage{longtable}

\usepackage{mathptmx}

\usepackage{amsmath}

\usepackage{amssymb}

\usepackage{graphicx}

\begin{document}

\begin{frontmatter}

\journal{Atomic Data and Nuclear Data Tables}

\copyrightholder{Elsevier Science}

\runtitle{Insert the running title here}
\runauthor{Insert running author names}


\title{Calculations of Maxwellian-averaged Cross Sections and Astrophysical Reaction Rates Using the ENDF/B-VII.0, JEFF-3.1, JENDL-3.3 and ENDF/B-VI.8 Evaluated Nuclear Reaction Data Libraries}

\author{B. Pritychenko}
\ead{pritychenko@bnl.gov}
\author{, S.F. Mughaghab, \MakeLowercase{and}}
\author{A.A. Sonzogni}
\address{National Nuclear Data Center, Brookhaven National Laboratory
\\Upton, NY 11973-5000, U.S.A.}

\begin{abstract}  
We calculated the Maxwellian-averaged cross sections (MACS) and astrophysical reaction rates of the stellar nucleosynthesis reactions (n,$\gamma$), (n,fission), (n,p), (n,$\alpha$) and (n,2n) using the ENDF/B-VII.0-, JEFF-3.1-, JENDL-3.3-, and ENDF/B-VI.8-evaluated nuclear-data libraries. Four major nuclear reaction libraries were processed under the same conditions for Maxwellian temperatures ({\it kT}) ranging from 1 keV to 1 MeV. We compare our current calculations of the {\it s}-process nucleosynthesis nuclei with previous data sets and discuss the differences between them and the implications for nuclear astrophysics. 
\end{abstract}

\end{frontmatter}





\newpage
\tableofcontents
\listofDtables

\vskip5pc

\section{Introduction}
Nuclear astrophysics emerged in the second half of the twentieth century as a frontier research field using nuclear-physics methods to resolve astrophysical problems \cite{88Rol}. One of the first successful results was an explanation by William Alfred Fowler of stellar nucleosynthesis. He was awarded the 1983 Nobel prize `` ... for his theoretical and experimental studies of the nuclear reactions of importance in the formation of the chemical elements in the universe'' \cite{83Fow}. 

Nuclear reactions play an important role in stellar nucleosynthesis and are responsible for the producing heavy chemical elements from light elements that were generated in the Big Bang. Heavy chemical elements are produced in the stars by different mechanisms, such as {\it s-} and {\it r-}processes \cite{88Rol}. Slow neutron capture (the {\it s}-process) that was found to operate in red giant stars explains the origin of about 60\% of the nuclei heavier than $^{56}$Fe \cite{08Sch}. The remaining 40\% of nuclei often are attributed to a rapid neutron-capture process, the {\it r}-process \cite{08Sch,04Cow}. In this work, we concentrate on the neutron-induced reactions responsible for the {\it s-}process. 

Present-day calculations of {\it s-}process nucleosynthesis calculations often are based on the dedicated nuclear astrophysics data tables, such as works of Bao {\it et al.} \cite{00Bao}, and Rauscher and Thielemann \cite{00Rau}. These data tables contain quality information on Maxwellian-averaged cross sections (MACS) and astrophysical reaction rates. However, it is essential to produce complementary neutron-induced reaction data sets for an independent verification and to expand the boundaries of the data tables. Recent releases of ENDF/B-VII.0-, JEFF-3.1-, JENDL-3.3-, and ENDF/B-VI.8-evaluated nuclear reaction libraries \cite{06Chad,02Jac,04Kon,02Shi,01Cse} and publication of the {\it Atlas of Neutron Resonances} reference book \cite{06Mugh} created a unique opportunity of applying these data for non-traditional applications, such as {\it s}-process nucleosynthesis \cite{88Rol,83Fow,00Bao,88Win}.

Previously, Nakagawa {\it et al.} \cite{05Nak} calculated MACS and astrophysical reaction rates from the JENDL-3.3-evaluated nuclear reaction library. In many cases, their results agree with the values recommended by Bao {\it et al.} \cite{00Bao} for the {\it s}-process nucleosynthesis neutron cross sections  while, in other cases, there are serious discrepancies. These divergences might originate from the problems with certain (n,$\gamma$) cross section evaluations in the JENDL-3.3 library \cite{02Shi}, in Bao {\it et al.'s} data sets \cite{00Bao} or in issues with the JENDL-3.3 calculations.

To resolve these discrepancies and exclude any problems with JENDL-3.3 processing as a possible reason, MACS and astrophysical reaction rates for ENDF/B-VII.0-, JEFF-3.1-, JENDL-3.3-, and ENDF/B-VI.8-evaluated nuclear reaction libraries were calculated at the National Nuclear Data Center (NNDC). A new Java-based code was developed for this purpose, and four major nuclear-reaction libraries were processed under the same conditions. We compare these results with the previous work in Nakagawa {\it et al.} \cite{05Nak} that was based on ENDF utility codes \cite{EUtil}, Bao {\it et al.} \cite{00Bao}, KADONIS database \cite{06Dil}, Rauscher and Thielemann \cite{00Rau}, the {\it Atlas of Neutron Resonances} reference book ({\it Atlas}) \cite{06Mugh}, and the recent calculations of MACS by Mughabghab \cite{08Mugh}. We describe the current work extensively in the following sections.

\section{Evaluated Nuclear Reaction Data Libraries}

The Evaluated Nuclear Data File (ENDF) is a core nuclear reaction library containing evaluated (recommended)
cross sections, spectra, angular distributions, fission-product yields,
thermal neutron scattering, photo-atomic, and other data, with an 
emphasis on neutron-induced reactions relevant to reactor calculations. In the United States, ENDF values are obtained via the following nuclear reaction codes and evaluation techniques \cite{06Chad}:
\begin{itemize}
 \item Hauser-Feshbach and pre-equilibrium, direct, and fission models for use in modeling medium-and-heavy-nucleus reactions, such as the GNASH \cite{96Yo} and COH \cite{06Ka} codes, and the EMPIRE \cite{07emp} code, which often is used in conjunction with coupled-channels optical-model codes such as ECIS \cite{94Ra}.
 \item R-matrix codes for light nucleus reactions, and for low-incident-energy reactions on heavier targets, 
 Los Alamos EDA \cite{81Ha}, and Oak Ridge SAMMY \cite{06La} codes.
 \item The {\it Atlas} code system \cite{06Mugh,00Oh} for analyzing neutron resonances in terms of Breit-Wigner formalism.
\end{itemize}

All evaluated nuclear-reaction data are stored in the
internationally adopted ENDF-6 format \cite{05He} maintained by the the Cross Section Evaluation Working Group (CSEWG) \cite{CSEWG}. Evaluation and dissemination of nuclear-reaction data are coordinated by the CSEWG  \cite{CSEWG} 
and the U.S. Nuclear Data Program (USNDP) \cite{USNDP} in the 
United States and by the Working Party on International Nuclear Data Evaluation Co-operation (WPEC) \cite{WPEC}.
 
A broad international effort led to the development of ENDF-6 formatted libraries: the ENDF in the United States \cite{06Chad,01Cse}, JEFF in Europe \cite{02Jac,04Kon}, JENDL in Japan \cite{02Shi}, BROND in Russia \cite{BROND},  and CENDL in China \cite{CENDL}. The United States, the European Union, and Japan continue to invest in research and development on evaluated nuclear-reaction libraries. This work resulted in the recent releases of the ENDF/B-VII.0 (U.S. 2006), JEFF-3.1 (Europe 2005), JENDL-3.3 (Japan 2002), and ENDF/B-VI.8 (U.S. 2001) libraries. Consequently, these four libraries, which presumably contain the most up-to-date nuclear data, were selected as a primary data source for our calculations of the {\it s}-process nucleosynthesis.

The ENDF libraries are based on theoretical calculations that often are adjusted to fit experimental data. There are two limiting situations in neutron-reaction physics: Direct reactions, and compound nucleus formation. Fig. \ref{fig0} shows an example of evaluated and experimental neutron-capture cross sections for $^{122}$Te. The experimental data are taken from the EXFOR/CSISRS database \cite{exfor} and evaluated data from the ENDF/B-VII.0 library \cite{06Chad}. The figure clearly demonstrates the importance of correctly estimating the contribution of neutron resonances in the astrophysically important keV-energy range.
\setcounter{figure}{0}

\begin{figure}
\begin{center}
\includegraphics[height=7cm]{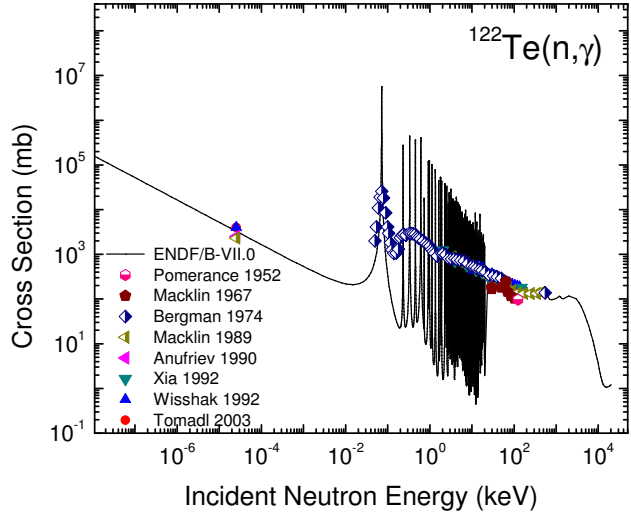}
\caption{ENDF/B-VII.0 evaluated and experimental neutron capture cross sections for $^{122}$Te \cite{52Po,67Mac,74Be,89Mac,90An,92Xi,92Wi,03To}.}
\label{fig0}
\end{center}
\end{figure}

\section{Calculation of Maxwellian-averaged Cross Sections, and Astrophysical Reaction Rates}

The ENDF-6 formatted libraries contain various data for all known nuclei,  including neutron capture cross sections  for 390 individual nuclei from $^{1}$H to $^{255}$Fm,  and a substantial number of elemental evaluations in the range of neutron energy from 10$^{-5}$ eV to 20 MeV. Nuclear-reactor and national-security application communities have used these data extensively in the eV and MeV energy ranges. However, there has been less exploration of data in the keV-energy-range that is of interest to the fast-reactor and nuclear-astrophysics communities.
   
\subsection{Processing Data from Evaluated Nuclear-Reaction Libraries}
Previously, Nakagawa {\it et al.} \cite{05Nak} calculated MACS and astrophysical reaction rates from the JENDL-3.3 library \cite{02Shi} using the ENDF utility codes \cite{EUtil}. In many cases, their resulting data agree well with other information \cite{00Bao,00Rau}, but, in some cases, serious discrepancies were observed. To clarify this issue, we undertook new nuclear-astrophysics calculations based on all evaluated nuclear reaction data \cite{06Chad,02Jac,04Kon,02Shi,01Cse,06Mugh} and the latest computer technologies.
 
The average cross sections for Maxwellian spectrum are defined as 
\begin{equation}
\label{myeq.a1}
\sigma^{Maxw}(kT) =  \frac{\langle \sigma \upsilon \rangle}{\upsilon_T}
\end{equation}
where $\upsilon$ is the relative velocity of neutrons and a target nuclide, and $\upsilon_{T}$ is the mean thermal velocity given by 
\begin{equation}
\label{myeq.a2}
\upsilon_{T} = \sqrt{\frac{2kT}{\mu}}
\end{equation}
where $\mu$ is the reduced mass of the target nucleus and neutron. The Maxwellian-averaged cross sections (MACS) can be expressed as   
\begin{equation}
\label{myeq.a3}
\sigma^{Maxw}(kT) = \frac{2}{\sqrt{\pi}}(kT)^{-2} \int_{0}^{\infty} \sigma(E)E e^{(- \frac{E}{kT})} dE
\end{equation}
where {\it k} and {\it T}, respectively,  are the Boltzman constant and temperature of the system,  and $E$ is an energy of relative motion of the neutron with respect to the target.  The values of the ENDF-6 formatted  cross sections  are tabulated in the laboratory system, and the last equation has to be modified to
\begin{equation}
\label{myeq.a4}
\sigma^{Maxw}(kT) = \frac{2}{\sqrt{\pi}} \frac{(m_2/(m_1 + m_2))^{2}}{(kT)^{2}}  \int_{0}^{\infty} \sigma(E^{L}_{n})E^{L}_{n} e^{- \frac{aE^{L}_{n}}{kT}} dE^{L}_{n}
\end{equation}
where  $E^{L}_{n}$ is a neutron energy in the laboratory system, and $m_{1}$ and $m_{2}$ are, respectively, the masses of a neutron and a target nucleus.

The astrophysical reaction rate, $R$, is defined as $R$ = $N_{A}$$\langle \sigma \upsilon \rangle$, where $N_{A}$ is the Avogadro number. To express reaction rates in [$cm^{3}$/mole s] units, an additional factor of $10^{-24}$ is introduced; $\upsilon_{T}$ is in units of [cm/s] and temperature, $kT$, in units of energy ({\it e.g.} MeV) is related to that in Kelvin ({\it e.g.} 10$^{9}$ K) as $T_{9}$=11.6045$kT$.
\begin{equation}
\label{myeq.a6}
R(T_{9}) = 10^{-24}N_{A}\sigma^{Maxw}(kT)\upsilon_{T}
\end{equation}

\begin{figure}
\begin{center}
\includegraphics[height=7cm]{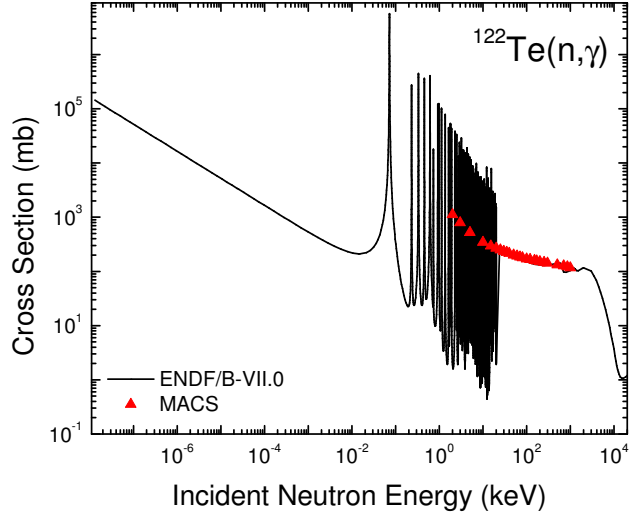}
\caption{ ENDF/B-VII.0-evaluated neutron capture cross sections and calculated MACS for $^{122}$Te.}
\label{fig1}
\end{center}
\end{figure}

Using these equations, we calculated MACS and astrophysical reaction rates based on the original evaluated neutron data from the ENDF libraries in the energy range from 1 keV to 1 MeV. Fig. \ref{fig1} demonstrates ENDF/B-VII.0-evaluated neutron capture cross sections, and calculated MACS for $^{122}$Te.

In the present work, the ENDF-6 formatted input files and mathematical formalism are similar to those of Nakagawa {\it et al.} \cite{05Nak}, while a new Java 1.6 ENDF library processing code was written to independently verify the results. The Java processing code was based on the Simpson rule of numerical integration and Java database connectivity to the Sigma Web Interface \cite{08Sig}.  The content of the Sigma database was Doppler-broadened and linearized via the PREPRO code \cite{00Cu} to serve as an input for numerical integration. We set the maximum width of the numerical integration interval (energy difference between evaluated cross section points) to less or equal to 10 keV. When the integration interval in the original ENDF library was more than 10 keV, additional evaluated cross sections were produced to satisfy this requirement. Further reduction of the maximal integration width to 1 keV did not affect the final result.  
 
\subsection{JENDL-3.3 Benchmarking Procedure}

\begin{figure}
\begin{center}
\includegraphics[height=8cm]{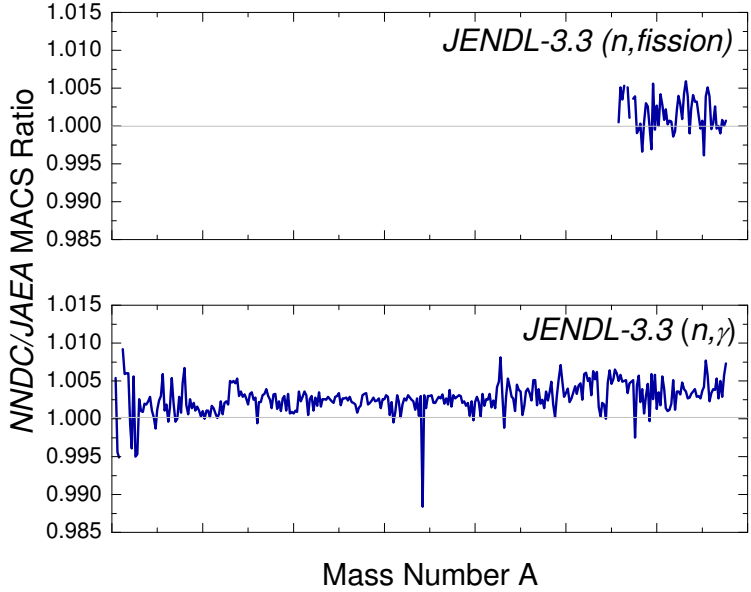}
\caption{MACS ratios between the NNDC  and the JAEA calculations for  neutron fission and capture reactions.}
\label{fig2}
\end{center}
\end{figure}

For benchmarking, we compared the MACS and astrophysical reaction rates from our current work with the Japan Atomic Energy Agency's (JAEA) JENDL-3.3 calculations \cite{05Nak}. Fig. \ref{fig2} illustrates the MACS ratios between the NNDC and the JAEA calculations for  neutron fission and capture at {\it kT}=30 keV. Both calculations for the JENDL-3.3 library agree within 0.1-0.9\%; the small differences between them reflects two different integration methods \cite{07Nak}.

\subsection{Tests of Nuclear-Astrophysics Data}

To verify the quality of ENDF libraries we tested several nuclear astrophysics data sets for {\it s}-process nuclei. 

\begin{figure}
\begin{center}
\includegraphics[height=7cm]{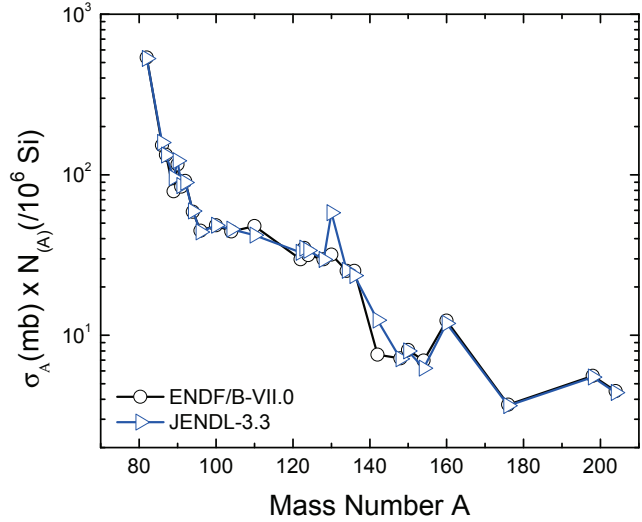}
\caption{ENDF/B-VII.0 and JENDL-3.3 $\langle \sigma^{Maxw}_{\gamma} (30 keV) \rangle$ multiplied by solar-system abundances (relative to Si = 10$^{6}$) are shown as a function of the mass number of nuclei produced in the {\it s}-process only.}
\label{fig4}
\end{center}
\end{figure}

First, the product of MACS and solar-system abundances ($N_{(A)}$) relative to Si is
\begin{equation} 
\label{myeq.s0} 
\sigma_{A}N_{(A)}= \sigma_{A-1}N_{(A-1)} = constant
\end{equation} 
for the equilibrium {\it s}-process-only nuclei \cite{88Rol,83Alm}. To verify this phenomenon, the calculated $\langle \sigma^{Maxw}_{\gamma} (30 keV) \rangle$ from the ENDF/B-VII.0 and JENDL-3.3 libraries were multiplied by solar abundances taken from Anders and  Grevesse \cite{89And}, and plotted  in Fig. \ref{fig4}; it reveals that the ENDF/B-VII.0 library data closely reproduces a two-plateau plot \cite{88Rol,83Alm,89Ka}, while JENDL-3.3 plot is off for two nuclei, $^{130}$Xe and $^{142}$Nd. These results indicate that the quality of nuclear-reaction data may explain the discrepancies between the calculation of Nakagawa {\it et al.} \cite{05Nak} and recommended values of Bao {\it et al.} \cite{00Bao}. For several nuclei, the nuclear-reaction data in the keV region of energies are based on theoretical models because  measurements with these nuclei are very difficult or lacking. Nuclear-astrophysics applications likely will initiate more measurements and eventual improvements of the ENDF evaluations over a wide range of energies and isotopes.

Second, the $\sigma_{A}N_{(A)}$ products ratio between  $^{122}$Te, $^{123}$Te, and $^{124}$Te that are shielded from the {\it r}-process by the isotopes of tin and antimony is expected to be 1:1:1  \cite{88Rol,82Kap}.  As listed in Table \ref{table01}, the results indicate that the {\it Atlas} data \cite{06Mugh} can serve as a benchmark for tellurium isotopes, while the evaluated nuclear-data library data for the $^{122,123,124}$Te isotopes  need improving. 
\begin{table}[hb]
\centering
\caption{ $\sigma_{A}N_{(A)}$ product ratios for neutron capture in  $^{122,123,124}$Te isotopes.}
\begin{tabular}{|l|c|} \hline
Data Source & Ratio of $^{122,123,124}$Te Products\\
\hline
ENDF/B-VII.0 \cite{06Chad} & 1.0:1.185:1.062\\
JEFF-3.1 \cite{04Kon} & 1.0:0.573:1.383\\
JENDL-3.3 \cite{02Shi} & 1.0:1.063:1.029\\
ENDF/B-VI.8 \cite{01Cse} & 1.0:0.573:1.416\\
{\it Atlas of Neutron Resonances} \cite{06Mugh} & 1.0:1.01:1.016\\
Bao {\it et al.} \cite{00Bao} & 1.0:0.973:0.97\\
\hline
\end{tabular} 
\label{table01}
\end{table}  

\begin{figure}[ht!]
\begin{center}
\includegraphics[height=8cm]{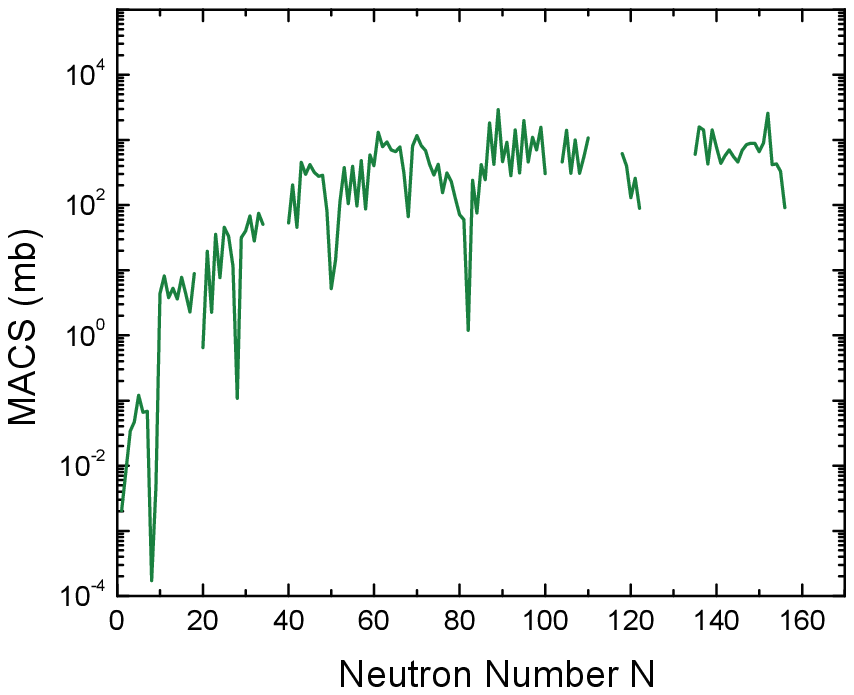}
\caption{MACS from ENDF/B-VII.0 library at {\it kT}=30 keV as a function of neutron number.}
\label{fig5}
\end{center}
\end{figure}

Third, trends in the MACS values should follow those of general nuclear physics. Fig. \ref{fig5} shows ENDF/B-VII.0   $\langle \sigma^{Maxw}_{\gamma} (30 keV) \rangle$ as a function of the neutron number, N. Large dips clearly are visible near the neutron-closed shells at N = 8, 20, 28, 50, 82, and 126. 

\section{Complementary Calculations of Maxwellian-averaged Cross Sections}

In addition to data on MACS at {\it kT=}30 keV ($\langle \sigma^{Maxw}_{\gamma} (30 keV) \rangle$)  from the ENDF libraries, {\it Atlas} \cite{06Mugh}, Bao {\it et al.} \cite{00Bao}, and KADONIS data \cite{06Dil},  Table \ref{Bao} includes complementary cross sections calculated by Mughabghab \cite{08Mugh}. These cross sections were produced to independently  verify the data, and are calculated from neutron-resonance parameters and nuclear systematics \cite{06Mugh,08bMugh,MughBNL}. 

Mughabghab \cite{08Mugh} followed two approaches, termed the detailed cross section method and the comparative method,  to determine $\langle \sigma^{Maxw}_{\gamma} (30 keV) \rangle$.

\subsection{The Detailed Cross Section Method}
In this procedure, the energy-dependent cross section is divided into two energy regions: 
\begin{itemize}
\item The thermal- and resolved-resonance region, accounted for by the individual resonance parameters in  {\it Atlas} \cite{06Mugh}.
\item The unresolved-resonance region, described by the average resonance parameters as reported in {\it Atlas} \cite{06Mugh} and determined in recent investigation \cite{08bMugh,MughBNL}. 
\end{itemize}
The latter region extends  up to 500 keV from the upper energy of the resolved-resonance region. The point-wise capture cross section then was computed by an in-house computer code RECENT \cite{07emp,00Cu}. Subsequently, the $\langle \sigma^{Maxw}_{\gamma} (30 keV) \rangle$ was determined by the INTER code \cite{inter}. More details are given by Oh {\it et al.} \cite{00Oh}.

Should the average resonance parameters not have been reported, these quantities are estimated via systematics described in the graphics of the {\it Atlas} \cite{06Mugh}, as well as by the recently determined trend for the ratio of average $p$-wave capture width, $\Gamma_{{\gamma}1}$, to that of the $s$-wave radiative width, $\Gamma_{{\gamma}0}$ \cite{08bMugh}.

\subsection{The Comparative Method}
This method is applied to those nuclei, $x$, above $Z=41$ for which information on the resonance parameters is not available, but MACS, as well as the gamma-strength functions, are well determined for other surrounding nuclei, $s$. For those $x$ and $s$ nuclei, the dominant contribution for the $\langle \sigma^{Maxw}_{\gamma} (30 keV) \rangle$ comes from the unresolved-resonance region. Thus, we can assess the capture cross section for those $x$ nuclei through the following simple relation,
\begin{equation}
\label{myeq.s1}
\langle \sigma_{\gamma x} \rangle = \langle \sigma_{\gamma s} \rangle \times S_{\gamma x} / S_{\gamma s}
\end{equation}
where, $\langle \sigma_{\gamma x} \rangle$, $ \langle \sigma_{\gamma s} \rangle$ are, respectively, the 30-keV capture cross sections of the $x$ and $s$ nuclei, and $ S_{\gamma x}$, $ S_{\gamma s}$ are, correspondingly, the gamma-strength functions of nuclei $x$ and $s$. 

The gamma-strength function is defined as the ratio of average radiative width $\Gamma_{\gamma l}$ to the average level spacing, $D_{l}$, for a particular orbital angular momentum, $l$. These quantities are estimated for the $s$- and $p$-wave neutrons for $x$ nuclei on the basis of systematics \cite{06Mugh,08bMugh} and applying the Gilbert-Cameron relation \cite{65Gi}. 

For those nuclei near the peaks of the $s$- or $p$-wave strength functions, the gamma-strength functions for $s$- and $p$-wave neutrons are applied respectively; otherwise, a  mean value is appropriately weighted by the neutron-strength functions.

\section{Analysis of Maxwellian-averaged Cross Sections and Reaction Rates}

Due to space limitations, we show, in   Tables \ref{Bao} - \ref{MACSTable18}, the MACS and astrophysical reaction rates  for (n,$\gamma$), (n,fission) reactions at temperatures {\it kT} = 1, 5, 10, 15, 20, 25, 30, 35, 40, 50, 70, 100, 200, 500 and 1000 keV using the ENDF/B-VII.0, JEFF-3.1, JENDL-3.3, and ENDF/B-VI.8 evaluated libraries. Complete data sets, calculated for temperatures {\it kT} = 1, 2, 3, 5, 10, 15, 20, 25, 30, 35, 40, 50, 60, 70, 80, 100, 120, 150, 170, 200, 250, 300, 500, 700, 850, and 1000 keV, including (n,p), (n,$\alpha$), and (n,2n) reaction data are available at the NNDC website \cite{06Pri}. The MACS and reaction-rate values for the whole ENDF energy range from 10$^{-5}$ eV to 20 MeV can be calculated on-line with the NucRates Web application {\it (http://www.nndc.bnl.gov/astro)}.

We discuss below our extensive analysis of the calculated MACS and reaction rates and compare them with previous results. 
Only data published before February 2009 were analyzed. Unpublished data sets, such as the evolving KADONIS database values  ({\it http://www.kadonis.org} or {\it http://nuclear-astrophysics.fzk.de/kadonis/}) \cite{06Dil}, are included in the data analysis and presented in Table \ref{Bao}.

\subsection{Maxwellian-averaged Cross Sections}
Tables \ref{Bao} and \ref{MACSTable102} list the MACS and reaction rate (n,$\gamma$) data sets from the ENDF/B-VII.0, JEFF-3.1, JENDL-3.3, and ENDF/B-VI.8-evaluated nuclear reaction (neutron) libraries, with  $\langle \sigma^{Maxw}_{\gamma} (30 keV) \rangle$ from the {\it Atlas of Neutron Resonances} ({\it Atlas}) \cite{06Mugh} and  the MACS calculation by Mughabghab \cite{08Mugh}; they also contain the recommended values from Bao {\it et al.}\cite{00Bao} and the KADONIS database \cite{06Dil}. The (n,fission)-evaluated nuclear-reaction library data sets appear in Table \ref{MACSTable18}.

\subsubsection{(n,$\gamma$) Reactions}

Table \ref{Bao} presents  the MACS data at {\it kT} = 30 keV  ($\langle \sigma^{Maxw}_{\gamma} (30 keV) \rangle$) from the evaluated nuclear-reaction libraries \cite{06Chad,02Jac,04Kon,02Shi,01Cse}, {\it Atlas} \cite{06Mugh},  recommended values and the NON-SMOKER calculations re-normalized to nearby experimental data by Bao {\it et al.} \cite{00Bao}, the KADONIS database \cite{06Dil},  and the resonance parameters calculations and nuclear systematics of Mughabghab \cite{08Mugh}.  In most cases, the recommendations of Bao {\it et al.} \cite{00Bao}, and the evaluated nuclear-data libraries and {\it Atlas} \cite{06Mugh} agree reasonably well. However, whenever the evaluated data disagreed with Bao {\it et al.'s} values \cite{00Bao} we  examined all available data in detail and generated the comments given for Table \ref{Bao}; they are supported by Mughabghab's calculation of neutron-resonance parameters and his systematics estimates \cite{08Mugh}, together with experimental data \cite{exfor}. In cases where experimental information is scarce, and target materials are unstable, new measurements were suggested to clarify the situation. These measurements may involve surrogate reactions on a deuterium target when the projectile (neutron) and that target ({\it s-}process nucleus) both are radioactive \cite{08Sch}, or a neutron-activation fast-cycling technique \cite{94Be,96Me}.

\begin{figure}
\begin{center}
\includegraphics[height=6cm]{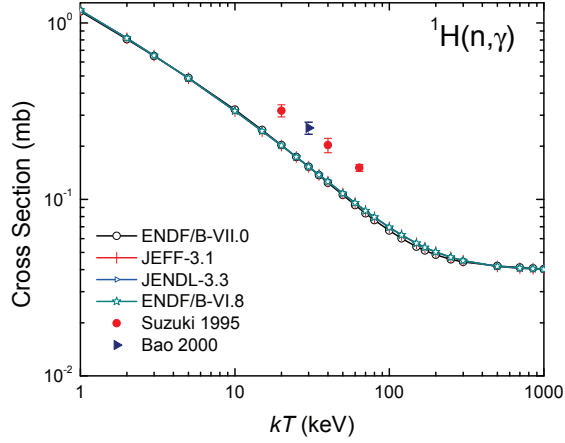}
\caption{Evaluated MACS and experimental cross sections for $^1$H \cite{00Bao,95Suz}.}
\label{1H}
\end{center}
\end{figure}

{\bf $^{1}$H}: Figure \ref{1H} demonstrates that the $\langle \sigma^{Maxw}_{\gamma} (30 keV) \rangle$  value of Bao {\it et al.} \cite{00Bao} is larger than the ones in the evaluated neutron libraries. Those authors adopted data from the fixed-target measurements of Suzuki {\it et al.} \cite{95Suz};  recalculating the cross section values with Eq. (\ref{myeq.a4}) without the $m_{2}/(m_{1} + m_{2})$ factor brings both results into agreement.

\begin{figure}
\begin{center}
\includegraphics[height=6cm]{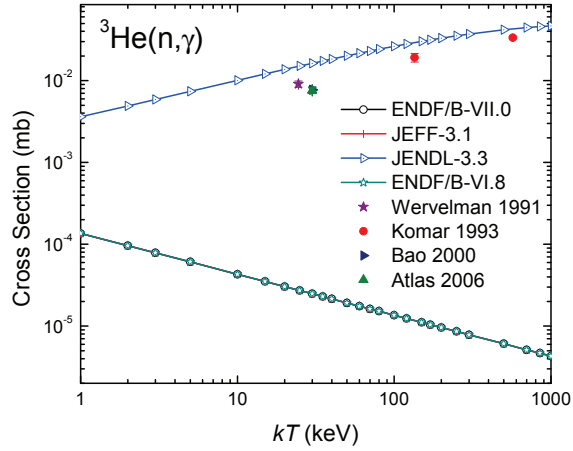}
\caption{Evaluated MACS and experimental cross sections for $^3$He \cite{00Bao,91Wer,93Ko}.}
\label{3He}
\end{center}
\end{figure}

{\bf $^{3}$He}: Visually inspecting Fig. \ref{3He} clearly verifies that the JENDL-3.3  MACS are consistent with the latest experimental results, while the ENDF/B-VII.0 library adopted old data from the ENDF/B-VI.8 library. The findings of Bao {\it et al.} \cite{00Bao} agree with those of {\it Atlas} $\langle \sigma^{Maxw}_{\gamma} (30 keV) \rangle$  \cite{06Mugh}, and both data sets are smaller by a factor of two than JENDL-3.3. This difference originates from the fact that Bao {\it et al.} \cite{00Bao} renormalized the data of Wervelman {\it et al.} \cite{91Wer} to the new standard of the $^{197}$Au(n,$\gamma$)$^{198}$Au reaction, whilst JENDL-3.3 uses these data without renormalization and consistent with many other data sets. Nakagawa {\it et al.} \cite{05Nak} offer an extensive review of $^{1}$H and $^{3}$He MACS. 

{\bf $^{10,11}$B}: Recent calculations of MACS by Mughabghab \cite{08Mugh} are, respectively, consistent with, and lower than ENDF libraries predictions for $^{11}$B and $^{10}$B. $^{10}$B calculated MACS agree well with Igashira {\it et al.'s} cross section measurements \cite{94Ig}.  

{\bf $^{12}$C}: The evaluated nuclear-reaction data libraries contain only elemental evaluation for carbon because the isotopic abundance of  $^{12}$C is 98.89\%. The JENDL-3.3 library's elemental evaluation correctly reproduces the latest experimental data \cite{91Na,92Oh,96Shi}, and agree with the values of Bao {\it et al.} \cite{00Bao} and {\it Atlas} \cite{06Mugh}  $\langle \sigma^{Maxw}_{\gamma} (30 keV) \rangle$ for $^{12}$C. The ENDF/B-VII.0, JEFF-3.1, and ENDF/B-VI.8 libraries use older data sets \cite{61Gib,90Mac} that entail lower cross section values.

\begin{figure}
\begin{center}
\includegraphics[height=6cm]{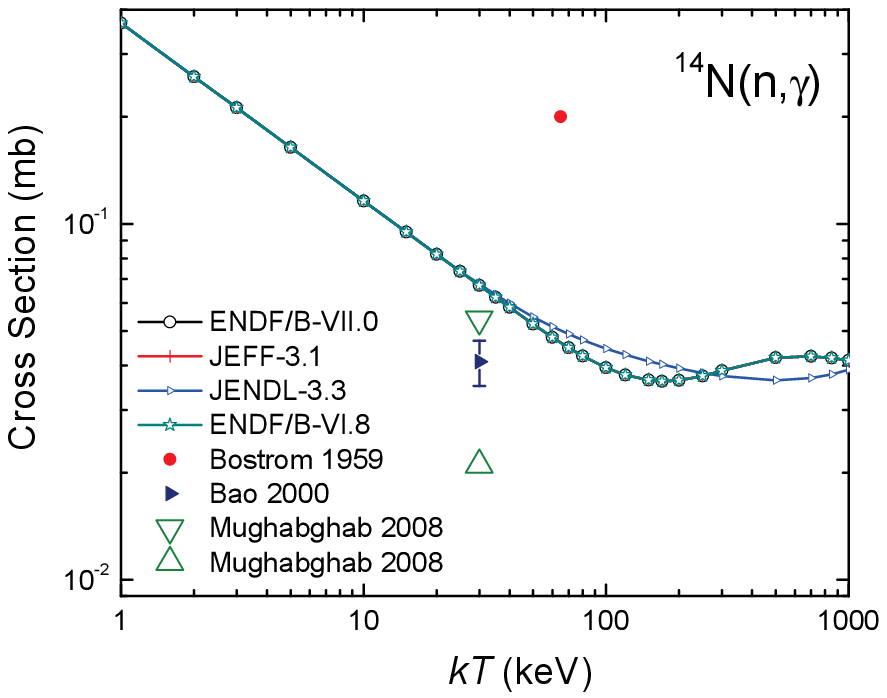}
\caption{Evaluated MACS and experimental cross sections for $^{14}$N \cite{00Bao,59Bo,08Mugh}.}
\label{14N}
\end{center}
\end{figure}

{\bf $^{14,15}$N}: Only the experimental cross sections of Bostrom {\it et al.} \cite{59Bo} and Meissner {\it et al.} \cite{96Me} are available for $^{14}$N and $^{15}$N, respectively.  The calculated  $\langle \sigma^{Maxw}_{\gamma} (30 keV) \rangle$ value is very sensitive to the bound-level position for $^{14}$N (Fig. \ref{14N}). This value is different from {\it Atlas} and BNL-325 data \cite{06Mugh,81BNL325} and eventually overlaps with Bao {\it et al.'s} recommendation \cite{00Bao}. The {\it Atlas} result for $^{15}$N \cite{06Mugh} is consistent with that of Bao {\it et al.} \cite{00Bao} and of Meissner {\it et al.} \cite{96Me}. Additional measurements in the keV energy range absolutely are needed for nitrogen isotopes.

\begin{figure}
\begin{center}
\includegraphics[height=6cm]{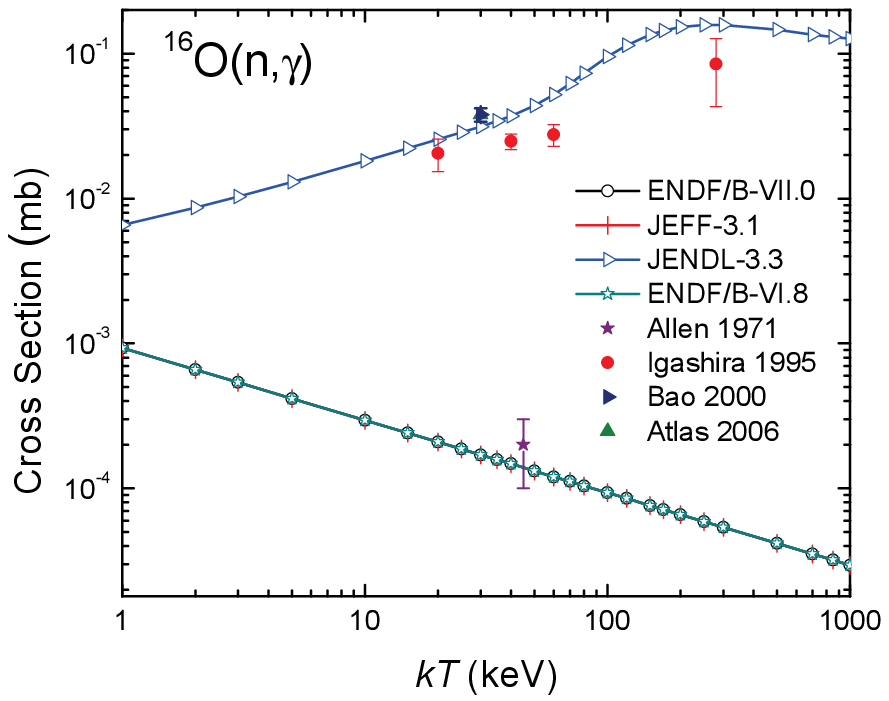}
\caption{Evaluated MACS and experimental cross sections for $^{16}$O \cite{00Bao,06Mugh,95Ig,71Al}.}
\label{16O}
\end{center}
\end{figure}

{\bf $^{16}$O}: Direct capture (the {\it p-}wave contribution) accounts for a significant fraction of the MACS in $^{16}$O \cite{00Bao}. Fig. \ref{16O} shows that JENDL-3.3 value accords with those of Bao {\it et al.} \cite{00Bao}, {\it Atlas} \cite{06Mugh} $\langle \sigma^{Maxw}_{\gamma} (30 keV) \rangle$,  and Igashira {\it et al.'s} recent experimental results \cite{95Ig}; the rest of the evaluated libraries rest on the older results of Allen and Macklin \cite{71Al}, and produce lower numbers.

{\bf $^{19}$F}:  The recent measurement of Uberseder {\it et al.} \cite{07Ube} is $\sim$44\%  lower than Bao {\it et al.'s} \cite{00Bao} cross section value. The latter value agrees with the 5.6$\pm$0.5 mb cross section calculated from the neutron-resonance parameters in {\it Atlas} \cite{06Mugh}.  

\begin{figure}
\begin{center}
\includegraphics[height=6cm]{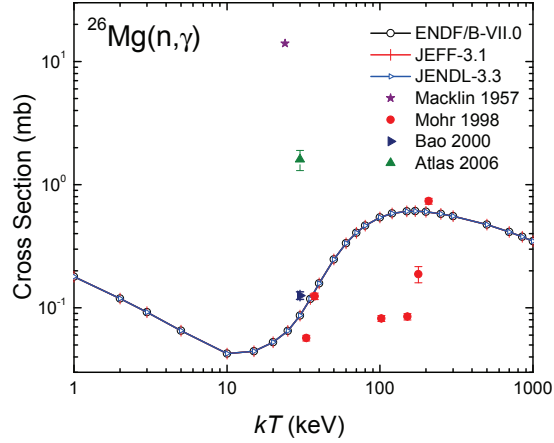}
\caption{Evaluated MACS and experimental cross sections for $^{26}$Mg \cite{00Bao,06Mugh,98Mo,57Mac}.}
\label{26Mg}
\end{center}
\end{figure}

{\bf $^{26}$Mg}: Here, contrary to the {\it s-}wave capture 1/$\upsilon$ cross-section trend, superposition of {\it s-} and {\it p-}wave components increases the cross section with the rise in neutron energy \cite{98Mo}. Fig. \ref{26Mg} demonstrates that most of $^{26}$Mg data sets are consistent, while {\it Atlas} \cite{06Mugh} affords an order-of-magnitude higher number that is influenced by the old finding of Macklin {\it et al.} \cite{57Mac}. 

{\bf $^{28}$Si}: MACS calculation for $^{28}$Si \cite{08Mugh} is consistent with the recent measurements of Guber {\it et al.} \cite{03Gu,02Gu}, 1.19$\pm$0.06 mb, providing evidence that the evaluated nuclear reaction libraries and Bao {\it et al.} \cite{00Bao} overestimate the  $\langle \sigma^{Maxw}_{\gamma} (30 keV) \rangle$  value.

{\bf $^{31}$P}: The JENDL-3.3 and JEFF-3.1 libraries closely reproduce the data of Macklin and Mughabghab \cite{85Mac} that was adopted by Bao {\it et al.} \cite{00Bao}; ENDF/B-VII.0 and ENDF/B-VI.8 data are based on the Macklin {\it et al.'s} older measurement \cite{63Mac}.

{\bf $^{33,36}$S}: There are few experimental  data sets available for $^{33,36}$S \cite{75Au,95Se}. $^{33}$S   $\langle \sigma^{Maxw}_{\gamma} (30 keV) \rangle$ calculated by Mughabghab \cite{08Mugh} agrees well with that of Bao {\it et al.} \cite{00Bao}, suggesting possible problems with ENDF libraries. The Bao {\it et al.} and {\it Atlas} data \cite{00Bao,06Mugh} correspond well with the experimental result of Sedyshev {\it et al.} \cite{95Se} for  $^{36}$S, giving much lower numbers than the evaluated nuclear libraries. However, more measurements are necessary for further clarification.

\begin{figure}
\begin{center}
\includegraphics[height=6cm]{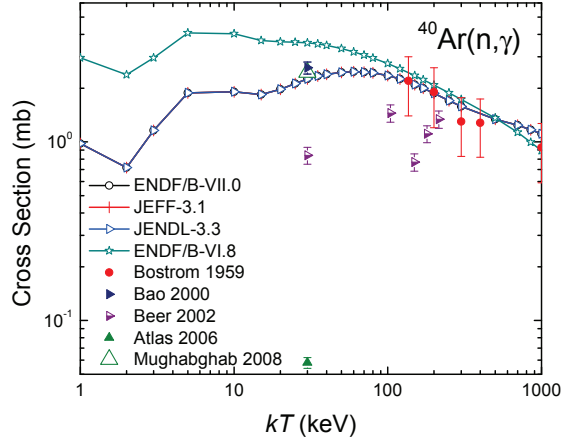}
\caption{Evaluated MACS and experimental cross sections for $^{40}$Ar \cite{00Bao,08Mugh,59Bo,02Be}.}
\label{40Ar}
\end{center}
\end{figure}

{\bf $^{38,40}$Ar}: Since experimental data are lacking for $^{38}$Ar, Bao {\it et al.} \cite{00Bao} adopted the theoretical $\langle \sigma^{Maxw}_{\gamma} (30 keV) \rangle$ of Rauscher and Thielemann \cite{00Rau}. This result is consistent with the nuclear systematics of Mughabghab \cite{08Mugh}, and more reliable than  evaluated nuclear-library values. For $^{40}$Ar shown in Fig. \ref{40Ar}, those libraries agree with Bao {\it et al.'s} calculations \cite{00Bao}, and the recent MACS assessment by Mughabghab \cite{08Mugh} while affording significantly higher values than Beer {\it et al.}  measurement \cite{02Be}. The $\langle \sigma^{Maxw}_{\gamma} (30 keV) \rangle$ value for $^{40}$Ar in the {\it Atlas of Neutron Resonances}  \cite{06Mugh} is low because of a typographical error.

\begin{figure}
\begin{center}
\includegraphics[height=6cm]{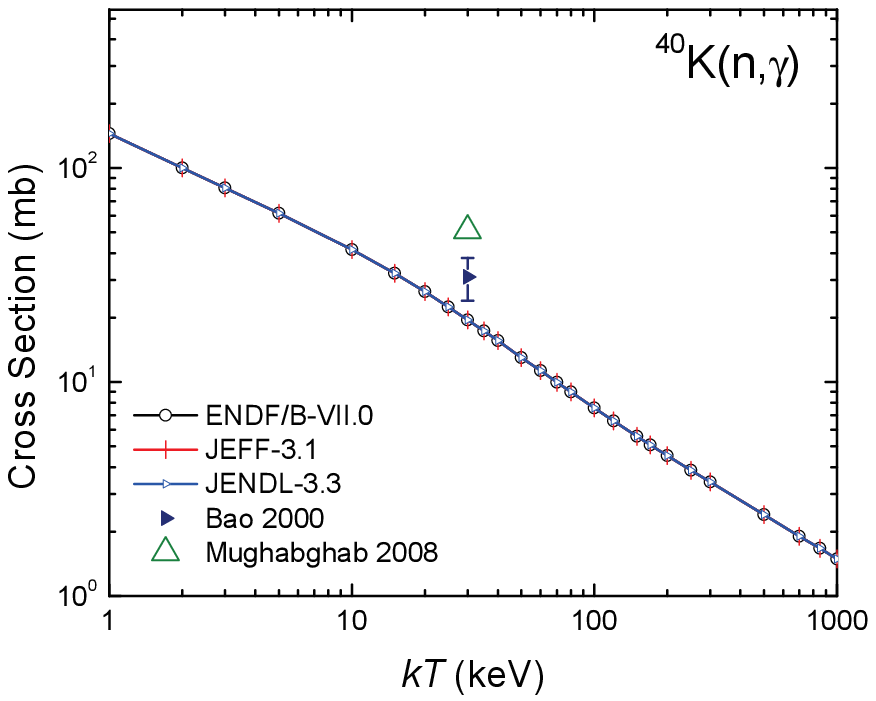}
\caption{Evaluated MACS for $^{40}$K \cite{00Bao,08Mugh}.}
\label{40K}
\end{center}
\end{figure}

{\bf $^{39,40,41}$K}: Bao {\it et al.} \cite{00Bao} adopted the 45 keV data of Macklin \cite{84Mac} for $^{39,41}$K that are in the agreement with the evaluated nuclear libraries.   The MACS for $^{40}$K are shown in Fig. \ref{40K}. Here, the ENDF/B-VII.0-adopted JENDL-3.3 evaluation, undertaken by H. Nakamura \cite{02Shi}, is based on CASTHY calculations \cite{75Ig}. Due to lack of experimental data for $^{40}$K, we cannot resolve the discrepancy between the neutron-resonance parameter calculation of Mughabghab \cite{08Mugh}, the values from Bao {\it et al.} \cite{00Bao}, and the evaluated nuclear libraries  $\langle \sigma^{Maxw}_{\gamma} (30 keV) \rangle$. $^{40}$K (T$_{1/2}$ = 1.248$\times$10$^{9}$ y) and $^{41}$K, partially produced from the $^{41}$Ca-$^{41}$K decay (T$_{1/2}$=1.02$\times$10$^{5}$ y), are important for cosmochronometry purposes \cite{95Was}.

{\bf $^{43}$Ca}: The evaluated nuclear-library cross sections are slightly lower than experimental results of Musgrove {\it et al.} \cite{77De}. At the same time, the figure given by Bao {\it et al.}  $\langle \sigma^{Maxw}_{\gamma} (30 keV) \rangle$ \cite{00Bao} agrees well with those of {\it Atlas} \cite{06Mugh}, the experimental cross sections \cite{exfor} and the recent calculation of Mughabghab \cite{08Mugh}.

\begin{figure}
\begin{center}
\includegraphics[height=6cm]{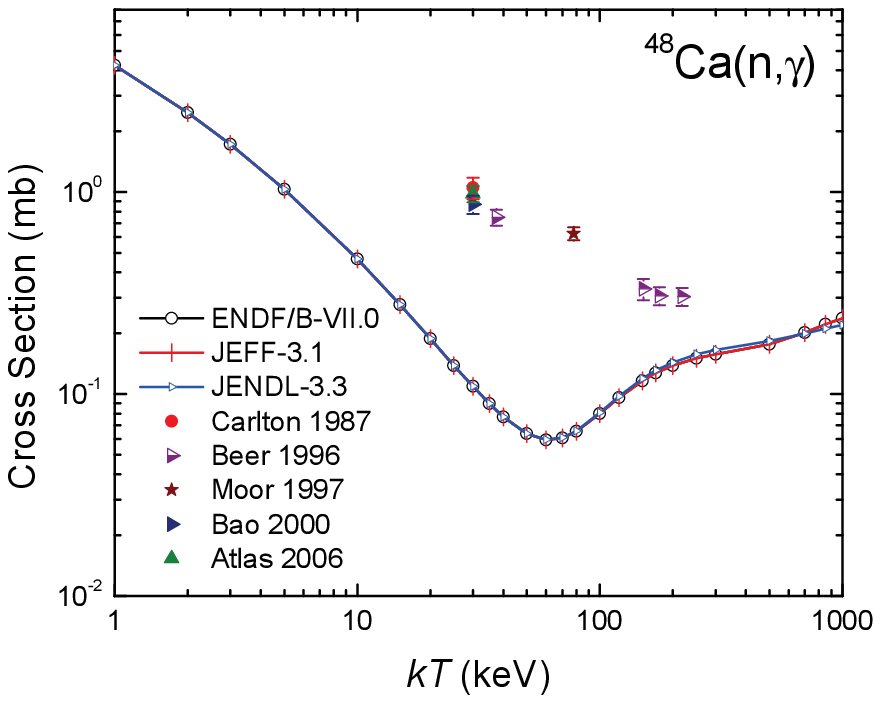}
\caption{Evaluated MACS and experimental cross sections for $^{48}$Ca \cite{00Bao,06Mugh,08Mugh,97Mo,96Bee,87Ca}.}
\label{48Ca}
\end{center}
\end{figure}

{\bf $^{46,48}$Ca}: The values of the evaluated nuclear-reaction library  $\langle \sigma^{Maxw}_{\gamma} (30 keV) \rangle$ are substantially lower than the experimental data for these isotopes. For the doubly magic nucleus of  $^{48}$Ca (T$_{1/2}$=4.3$\times$10$^{19}$ y), a problem is introduced by the negative-energy resonance \cite{87Ca} in the JEFF-3.1 evaluation, and the neglected contribution from the direct-capture mechanism, as shown in Fig. \ref{48Ca}. Consequently, Bao {\it et al.'s} values \cite{00Bao} are recommended for $^{46,48}$Ca. These issues with the JEFF-3.1 library's cross section values do not represent a major problem for nuclear industry-applications because the isotopic abundances of  $^{46}$Ca and $^{48}$Ca are, respectively,  0.004\% and 0.187\%.

\begin{figure}
\begin{center}
\includegraphics[height=6cm]{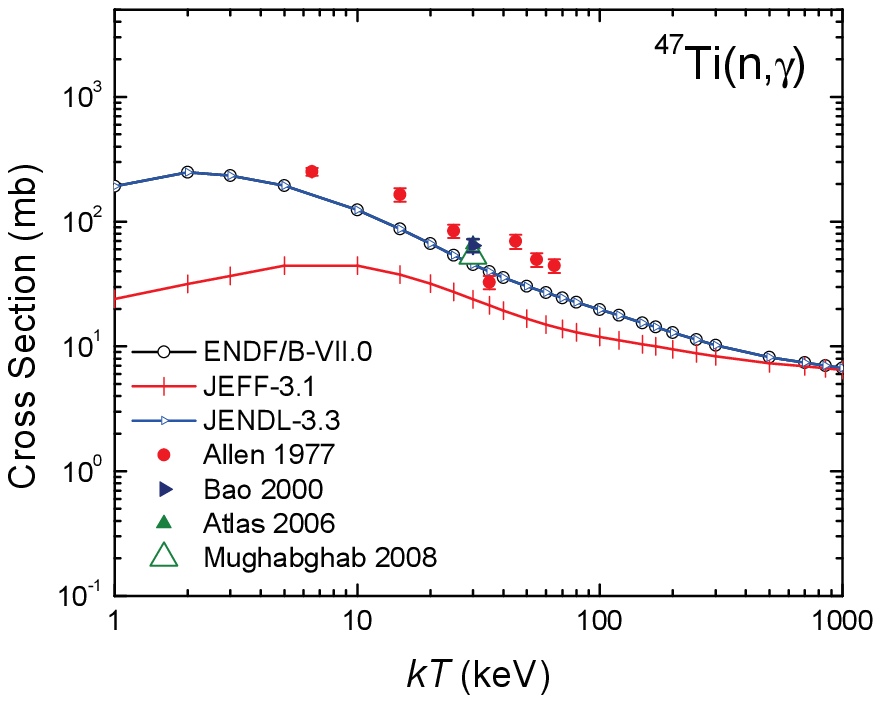}
\caption{Evaluated MACS and experimental cross sections for $^{47}$Ti \cite{00Bao,06Mugh,08Mugh,77Al}.}
\label{47Ti}
\end{center}
\end{figure}

{\bf $^{47}$Ti}: As shown in Fig. \ref{47Ti}, the figures given by Bao {\it et al.} and Mughabghab  $\langle \sigma^{Maxw}_{\gamma} (30 keV) \rangle$ for  $^{47}$Ti \cite{00Bao,06Mugh,08Mugh} agree with the experimental results of Allen {\it et al.} \cite{77Al}, and somewhat higher than those in evaluated libraries.

{\bf $^{51}$V}: Evaluated nuclear-data libraries contain only elemental evaluations for vanadium because isotopic abundance of $^{51}$V is 99.75\%. The JENDL-3.3 elemental  $\langle \sigma^{Maxw}_{\gamma} (30 keV) \rangle$ is very close to the adopted values of Bao {\it et al.} \cite{00Bao}, and Mughabghab \cite{06Mugh} for $^{51}$V. 

{\bf $^{53}$Cr}: The evaluated   $\langle \sigma^{Maxw}_{\gamma} (30 keV) \rangle$ is somewhat lower than adopted values of Bao {\it et al.} \cite{00Bao}, Mughabghab \cite{06Mugh}, and the experimental cross sections of Kenny {\it et al.} \cite{77Ke} and Beer {\it et al.} \cite{75Be}. The assessed $^{53}$Cr resonance parameters of Mughabghab \cite{08Mugh} are consistent with the evaluated nuclear libraries.

{\bf $^{55}$Mn}: The MACS value determined from resonance parameters \cite{06Mugh} is 32.2 mb, consistent with the evaluated nuclear libraries, but in contrast with  $\langle \sigma^{Maxw}_{\gamma} (30 keV) \rangle$ =39.6$\pm$3.0 mb, recalculated by Bao {\it et al.} \cite{00Bao} using the older version of the resonance parameters data \cite{81BNL325}. 

{\bf $^{54}$Fe}: The {\it Atlas} cross section value, which is  higher than those in the evaluated nuclear libraries, is based on data taken from  Allen {\it et al.} ( $\sigma_{\gamma}$=33.6$\pm$2.7 mb) \cite{Al77}. On the other hand, the calculated $\langle \sigma^{Maxw}_{\gamma} (30 keV) \rangle$, derived from the {\it Atlas} resonance parameters \cite{06Mugh}, yields a value of 27.4 mb that is in excellent agreement with Bao {\it et al.'s}  renormalization \cite{00Bao} of Brusegan {\it et al.}  result \cite{83Bru}, 27.6$\pm$1.8 mb.

{\bf $^{56}$Fe}: The {\it Atlas}  cross section value of 15.1$\pm$1.3 mb is attributed to measurements at ORNL \cite{76Al}. This figure differs substantially from the values generated from the data libraries, and that recommended by Bao {\it et al.} \cite{00Bao}. This discrepancy reflects the large capture width of 1.43 eV for the 27.79 keV resonance reported in the former publication. Later measurements showed that the capture width of this resonance is 1.0$\pm$0.1 eV \cite{81Wi,92Co}; accordingly, the reconstructed resonance parameters yield  $\langle \sigma^{Maxw}_{\gamma} (30 keV) \rangle$ of 10.8 mb.

{\bf $^{57}$Fe}: A Maxwellian  $\langle \sigma^{Maxw}_{\gamma} (30 keV) \rangle$ of 34.7$\pm$2.3 mb was reported by Allen {\it et al.} \cite{77All}, and a value of 37.7 mb is computed on  the basis of the {\it Atlas} parameters \cite{06Mugh}. We note that the inelastic channel for $^{57}$Fe(n,n$^{\prime}$) that opens at 14.5 keV is accounted for in our calculation. Bao {\it et al.'s} value of 40$\pm$4 mb \cite{00Bao} is based on the resonance parameters of Rohr {\it et al.} \cite{83Ro}.

\begin{figure}
\begin{center}
\includegraphics[height=6cm]{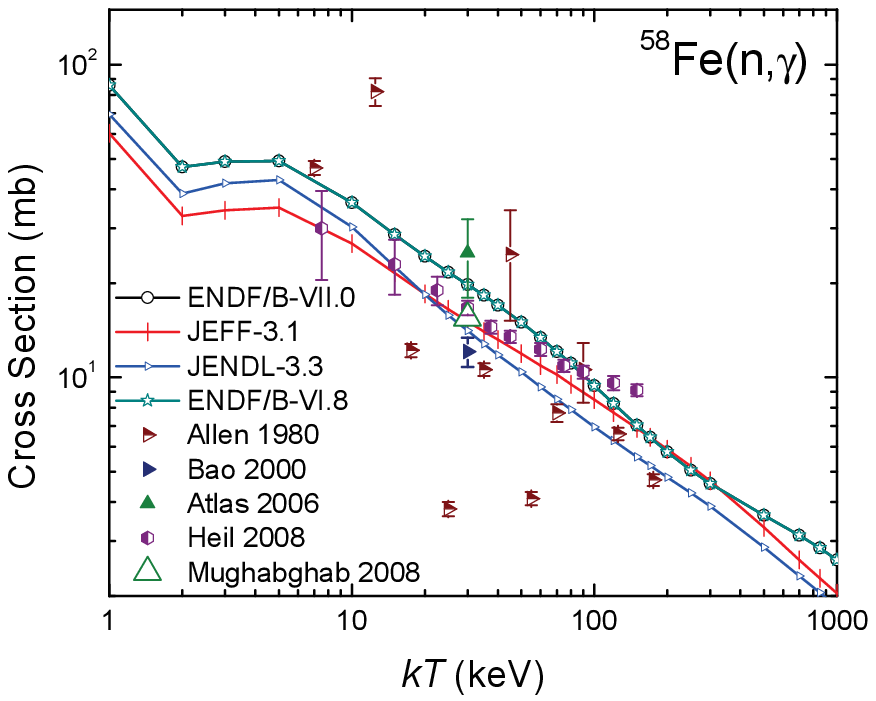}
\caption{Evaluated MACS and experimental cross sections for $^{58}$Fe \cite{00Bao,06Mugh,08Mugh,08He,80All}.}
\label{58Fe}
\end{center}
\end{figure}

{\bf $^{58}$Fe}: The relatively large value in {\it Atlas}  $\langle \sigma^{Maxw}_{\gamma} (30 keV) \rangle$ value of 25$\pm$7 mb \cite{06Mugh} rests on the measurements of Hong {\it et al.} (24.6$\pm$6.5 mb) \cite{77Ho}, while the calculation of resonance parameters predicts 15.8 mb \cite{08Mugh}. Fig. \ref{58Fe} demonstrates that this value corresponds to the recent result of Heil {\it et al.} \cite{08He}, and the evaluated libraries. Allen and Macklin report a value of 15.9$\pm$1.5 mb \cite{80All}. Later measurements at Forschungszentrum Karlsruhe, Institut f\"{u}r Kernphysik (KFK) with a better signal-to-background ratio yielded 14.3$\pm$1.4 mb \cite{83Kae}. The recommended value of Bao {\it et al.} (12.1$\pm$1.3 mb) \cite{00Bao} represents a renormalized KFK result. 

{\bf $^{62}$Ni}: The ENDF/B-VII.0 and JENDL-3.3 libraries provide MACS values of 51.7 mb and 19.2 mb, respectively. Bao {\it et al.} \cite{00Bao} recommend  a  $\langle \sigma^{Maxw}_{\gamma} (30 keV) \rangle$ of 12.5$\pm$4.0 mb. The source of the discrepancy between the ENDF libraries and the others is the inclusion by the libraries of a background capture cross section in the energy region from 10 to 600 keV. Based on {\it Atlas} \cite{06Mugh}, the resonance contribution to  $\langle \sigma^{Maxw}_{\gamma} (30 keV) \rangle$ is 13.59 mb. The direct thermal-capture cross section 1/$\upsilon$-extrapolation to 30 keV yields a contribution of 8.87 mb, resulting in the total cross section of 22.5 mb \cite{08Mugh}. This value better conforms with the 26$\pm$5 mb reported by Beer {\it et al.} \cite{74Bee}, and is consistent with the recent measurements of  a $^{62}$Ni(n,$\gamma$) cross section by Alpizar-Vicente {\it et al.} (25.8$\pm$2.6 mb) \cite{08Al} and Tomyo {\it et al.} (37.0$\pm$3.2 mb) \cite{05To}.

{\bf $^{64}$Ni}: There is a large discrepancy between the MACS generated using the ENDF libraries and that recommended by Bao {\it et al.} \cite{00Bao}, viz., 22.1 mb, and 8.7$\pm$0.9 mb, respectively. The latter rests on Heil {\it et al.} measurements \cite{08He}. We attribute the source of this difference to the large capture widths of the $s$-wave resonances located at 14.3 and 33.81 keV, $\Gamma_{\gamma}$=1.97 and 2.9 eV, respectively. {\it Atlas} recommends  capture widths of 1.01$\pm$0.07 and 1.16$\pm$0.08 eV \cite{06Mugh}, respectively, based on the  Wisshak {\it et al.'s} measurement at KFK \cite{84Wis}. The {\it Atlas} parameters reveal that the resonance contribution to  $\langle \sigma^{Maxw}_{\gamma} (30 keV) \rangle$ is 12.1$\pm$1.0 mb.  Furthermore, Mughabghab \cite{06Mugh} showed that 83.5\% of the thermal-capture cross section, 1.37 b, is due to direct capture. This component contributes 1.3 mb to the  $\langle \sigma^{Maxw}_{\gamma} (30 keV) \rangle$, resulting in a total value of 13.4$\pm$1.0 mb.

\begin{figure}
\begin{center}
\includegraphics[height=6cm]{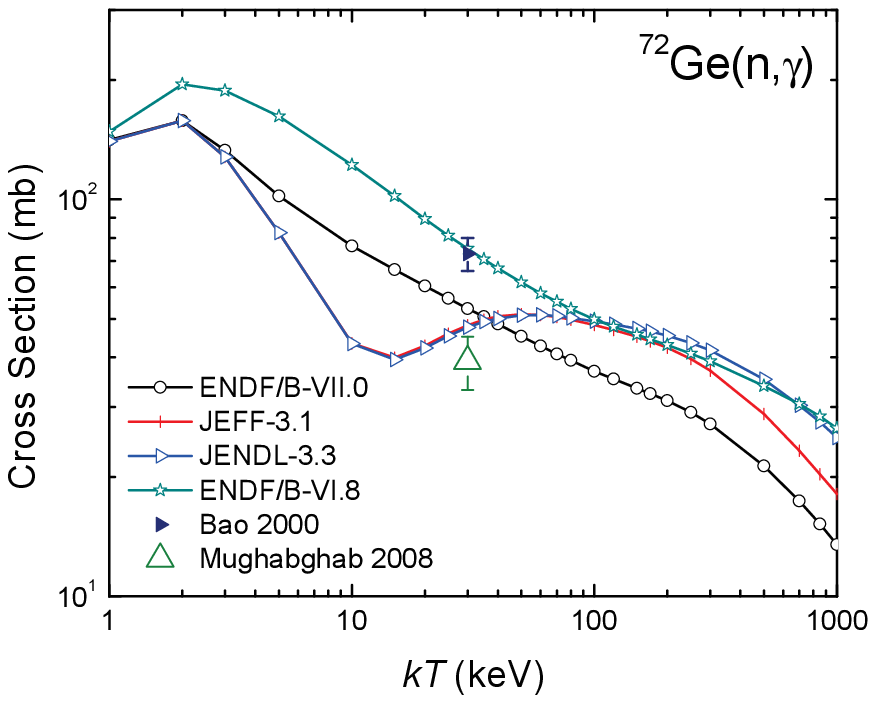}
\caption{Evaluated MACS for $^{72}$Ge \cite{00Bao,08Mugh}.}
\label{72Ge}
\end{center}
\end{figure}

{\bf $^{72}$Ge}: As shown in Fig. \ref{72Ge}, the ENDF/B-VII.0 library \cite{06Chad} and nuclear systematics of Bao {\it et al.} \cite{00Bao} give 30-keV MACS values of 53.1 mb and 73$\pm$7 mb, respectively. On the other hand, the {\it Atlas} resonance parameters \cite{06Mugh} along with the average resonance parameters, specify a value of 39.1$\pm$6.0 mb that is inconsistent, with Bao {\it et al.'s} \cite{00Bao} determination but close to the ENDF/B-VII.0 value.

{\bf $^{73}$Ge}: Using the nuclear systematics, Bao {\it et al.} \cite{00Bao} recommend $\langle \sigma^{Maxw}_{\gamma} (30 keV) \rangle$  of 243$\pm$47 mb, while the data libraries give values from 191 mb to 282 mb. Since the resolved-resonance parameters of $^{73}$Ge extend only to 8.9 keV, the major contribution to  $\langle \sigma^{Maxw}_{\gamma} (30 keV) \rangle$ originates in the unresolved-resonance region. The value calculated from the parameters of these two regions is 190$\pm$20 mb \cite{08Mugh}.

{\bf $^{76}$Ge} (T$_{1/2}$=1.3$\times$10$^{21}$ y): The {\it Atlas's} recommended  $\langle \sigma^{Maxw}_{\gamma} (30 keV) \rangle$ value, 16.0$\pm$1.6 mb, agrees well with ENDF/B-VII.0 library. On the other hand, the value calculated from the {\it Atlas} resolved- and unresolved-resonance parameters is 16.2$\pm$5.2 mb. The large uncertainty in this value primarily is due to the large uncertainty of the capture widths. In contrast, Bao {\it et al.'s} recommendation \cite{00Bao}, 33$\pm$15 mb, is based on systematics.

{\bf $^{74,76}$Se}:  There is agreement between the Bao {\it et al.},  and {\it Atlas} cross sections \cite{00Bao,06Mugh} and the recent measurement of Dillmann {\it et al.} \cite{06Di} for $^{74}$Se. The $^{76}$Se unresolved-resonance region, 9 keV - 100 keV, in both ENDF/B-VII.0 and JENDL-3.3 are based on CASTHY model calculations \cite{75Ig}, yielding  $\langle \sigma^{Maxw}_{\gamma} (30 keV) \rangle$ of 96 mb.  Beer {\it et al.} \cite{Be92} report a measurement of 164$\pm$8 mb that falls readily within the uncertainty limits of the calculation based on the {\it Atlas} parameters, and the recently determined systematics of capture widths \cite{08Mugh,08bMugh}, yielding  $\langle \sigma^{Maxw}_{\gamma} (30 keV) \rangle$ = 148$\pm$22 mb. It is interesting to note that the JEFF-3.1 value, 155 mb, is based on a $\sigma (n,\gamma)$ adopted from ENDF/B-VI.8.

{\bf $^{79}$Se} (T$_{1/2}$=2.95$\times$10$^{5}$ y): There is no resonance information from direct measurements for $^{79}$Se. Bao {\it et al.} recommend a    $\langle \sigma^{Maxw}_{\gamma} (30 keV) \rangle$ value of 263$\pm$46 mb \cite{00Bao} obtained from systematics while the data libraries generate  a $\langle \sigma^{Maxw}_{\gamma} (30 keV) \rangle$ of 416 mb. The average resonance parameters for $^{79}$Se were deduced from those of $^{77}$Se and recent systematics for $\Gamma_{\gamma_1}$/$\Gamma_{\gamma_0}$  \cite{08Mugh,08bMugh}. Calculations based on these resonance parameters give a  $\langle \sigma^{Maxw}_{\gamma} (30 keV) \rangle$ = 437$\pm$60 mb,  {\it i.e.}, inconsistent with Bao {\it et al.'s} recommendation \cite{00Bao} and agreeing well with the value derived from the evaluated libraries. 

\begin{figure}
\begin{center}
\includegraphics[height=6cm]{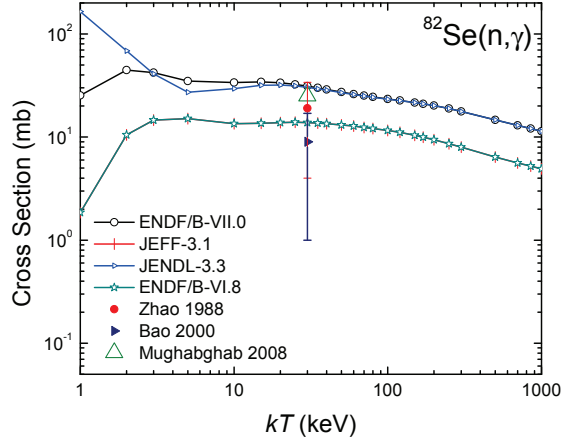}
\caption{Evaluated MACS and experimental cross sections for $^{82}$Se \cite{00Bao,08Mugh,88Zh}.}
\label{82Se}
\end{center}
\end{figure}

{\bf $^{82}$Se} (T$_{1/2}$=9.2$\times$10$^{19}$ y): Using systematics for $^{82}$Se, Bao {\it et al.} \cite{00Bao} estimate  a $\langle \sigma^{Maxw}_{\gamma} (30 keV) \rangle$ of 9$\pm$8 mb. In contrast, the ENDF/B-VII.0 and ENDF/B-VI.8 libraries report values of 31.8 and 13.8 mb, respectively. To resolve this inconsistency, we carried out calculations in terms of resolved- ($<$ 7 keV) and unresolved-resonance parameters (7 - 500 keV). One of the main uncertainties in this approach is attributed to the imprecise value of the {\it s-}wave level spacings, $D_{0}$ = 6.7$\pm$4.7 keV \cite{06Mugh}. Using the Gilbert-Cameron relation \cite{65Gi} and $D_{0}$ = 1.48$\pm$0.20 keV for $^{78}$Se, a value of 6.7 keV is confirmed for the level spacing of $^{82}$Se. With the average resonance parameters \cite{06Mugh}, as well as systematics \cite{08bMugh}, a $\langle \sigma^{Maxw}_{\gamma} (30 keV) \rangle$ of 25$\pm$7 mb is deduced for $^{82}$Se. Interestingly, Fig. \ref{82Se} reveals that this result is consistent with that of Zhao {\it et al.} \cite{88Zh} deduced from their derived systematics,  $\langle \sigma^{Maxw}_{\gamma} (30 keV) \rangle$ = 19$\pm$6 mb.

\begin{figure}
\begin{center}
\includegraphics[height=6cm]{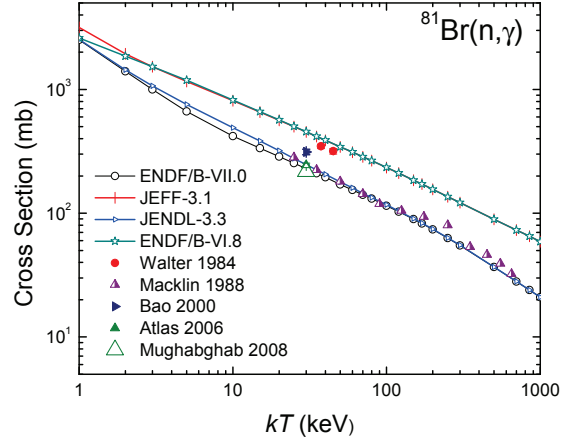}
\caption{Evaluated MACS and experimental cross sections for $^{81}$Br \cite{00Bao,06Mugh,08Mugh,86Wa,88Ma}.}
\label{81Br}
\end{center}
\end{figure}

{\bf $^{81}$Br}: The ENDF/B-VII.0 and JENDL-3.3 evaluated libraries yield  $\langle \sigma^{Maxw}_{\gamma} (30 keV) \rangle$ = 229 mb, and 247 mb, respectively,  that are inconsistent with Bao {\it et al.'s} recommendation \cite{00Bao}, 313$\pm$16 mb, as shown in Fig. \ref{81Br}. The latter value is supported by the measurements of Walter {\it et al.} \cite{86Wa}. However, the recommendation in {\it Atlas} \cite{06Mugh},  $\langle \sigma^{Maxw}_{\gamma} (30 keV) \rangle$ = 244$\pm$10 mb, is founded on the Macklin resonance parameters of the resolved- and unresolved-energy regions \cite{88Ma}. Not surprisingly, the JENDL-3.3 value, 247 mb, is close to the Macklin result because this number is normalized to the Macklin capture cross section in the unresolved-energy region. Adopting the well-determined average resonance parameters of {\it Atlas}, as well as an average capture width of 300 meV for $p$-wave resonances \cite{08bMugh}, Mughabghab \cite{08Mugh} calculates 288$\pm$19 mb for the $\langle \sigma^{Maxw}_{\gamma} (30 keV) \rangle$, compatible with Walter {\it et al.'s} result  \cite{86Wa}, within the uncertainty limits. The discrepancy may be due to water absorption in the ORNL sample, $NaBr-2H_{2}O$. We note that by considering the actual energy-dependence of the Maxwellian capture cross section \cite{08Mugh}, rather than assuming a $1/{\upsilon}$ dependence for 25-30 keV neutrons, the 313$\pm$16 mb becomes 309$\pm$16 mb.

\begin{figure}
\begin{center}
\includegraphics[height=6cm]{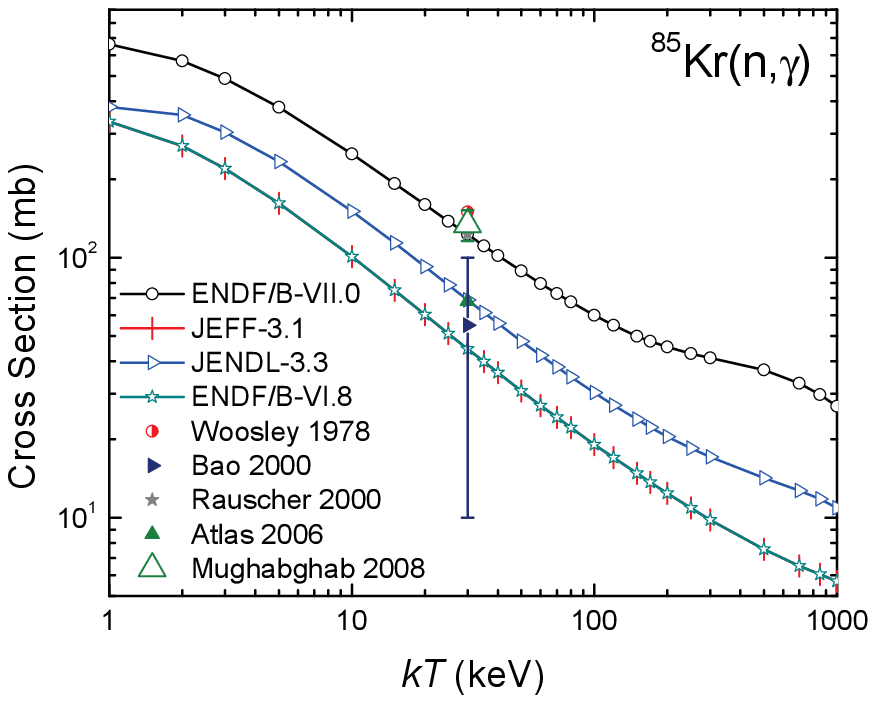}
\caption{Evaluated MACS for $^{85}$Kr \cite{00Bao,00Rau,06Mugh,08Mugh,80Le,78Wo}.}
\label{85Kr}
\end{center}
\end{figure}

{\bf $^{85}$Kr} (T$_{1/2}$ = 10.76 y): Fig. \ref{85Kr} indicates lack of experimental data for the radioactive isotope of $^{85}$Kr.   Bao {\it et al.} \cite{00Bao} renormalized the NON-SMOKER calculations to nearby experimental data, ( {\it i.e.} Bao {\it et al.} nuclear systematics \cite{00Bao}) with large errors. The {\it Atlas} value \cite{06Mugh}, 68 mb, rests on a calculation of Leugers {\it et al.} \cite{80Le}. The ENDF/B-VII.0 and JENDL-3.3 nuclear libraries produce  $\langle \sigma^{Maxw}_{\gamma} (30 keV) \rangle$ of 123 mb and 68.7 mb. The major source of the discrepancy is attributed to the widely different values of the average-level spacing adopted by these libraries. To resolve the discrepancies, Mughabghab \cite{08Mugh} calculated corresponding values of 152 mb and 116 mb following the detailed cross section and comparative methods, respectively. The latter estimate is based on adopting $^{85}$Rb as the comparative nucleus with   $\langle \sigma^{Maxw}_{\gamma} (30 keV) \rangle$=240$\pm$9 mb \cite{00Bao}. Aided by these results, an average, 134$\pm$18 mb, is obtained which is compatible with  Rauscher and Thielemann \cite{00Rau} and  and Woosley {\it et al.} \cite{78Wo} values, 123 mb and 150 mb, respectively.

{\bf $^{92}$Zr}: Although the findings of Boldemann {\it et al.} at ORNL \cite{76Bo}, $\langle \sigma^{Maxw}_{\gamma} (30 keV) \rangle$ = 33$\pm$4 mb, was later revised by Musgrove {\it et al.} \cite{78Mus} to $\langle \sigma^{Maxw}_{\gamma} (30 keV) \rangle$ = 50$\pm$6 mb; the former was recommended by Bao {\it et al.} \cite{00Bao} and Mughabghab \cite{06Mugh}. In terms of the {\it Atlas's} resonance parameters \cite{06Mugh}, the calculated value is 47.2$\pm$4.7 mb \cite{08Mugh}. The data libraries yield a result of 45.7 mb, conforming with the  recent measurement of the $^{92}$Zr(n,$\gamma$) reaction by Ohgama {\it et al.} \cite{05Oh}. 

{\bf $^{95}$Zr} (T$_{1/2}$ = 64.03 d): Since the half-life of $^{95}$Zr is comparatively  short, cross section data and resonance parameter information are unavailable. Consequently, Bao {\it et al.} \cite{00Bao} employed nuclear systematics to derive  $\langle \sigma^{Maxw}_{\gamma} (30 keV) \rangle$ = 79$\pm$12 mb. However, the ENDF/B-VII.0-, JEFF-3.1-, and JENDL-3.3 values generated from the capture cross sections are 140 mb, 51.3 mb and 140 mb, respectively. Mughabghab \cite{08Mugh} followed two approaches, the comparative method and the detailed method to calculate  $\langle \sigma^{Maxw}_{\gamma} (30 keV) \rangle$. In the former case, $\langle \sigma^{Maxw}_{\gamma} (30 keV) \rangle$ was computed by a scaling procedure whereby the known $^{93}$Zr   $\langle \sigma^{Maxw}_{\gamma} (30 keV) \rangle$ is scaled by the $p$-wave gamma-strength functions of $^{95}$Zr relative to $^{93}$Zr, so generating 51$\pm$14 mb. In the latter procedure, the average resonance parameters of $^{95}$Zr are derived via systematics, and then the capture cross section of $^{95}$Zr is computed by the code RECENT \cite{07emp,00Cu}, from which a  $\langle \sigma^{Maxw}_{\gamma} (30 keV) \rangle$=67$\pm$19 mb, is calculated. A weighted-average value, 57$\pm$11 mb, then is obtained.

{\bf $^{94}$Nb} (T$_{1/2}$ = 2.03$\times$10$^{4}$ y): Since the only resonance parameters available for $^{94}$Nb  are below 23 eV \cite{06Mugh}, the capture cross section for this isotope above this energy is obtained by nuclear-model calculations, along with dependence on the systematics of average resonance parameters. The evaluated data libraries give $\langle \sigma^{Maxw}_{\gamma} (30 keV) \rangle$ = 318 mb from the nuclear-model calculations, CASTHY \cite{75Ig}. With systematics, Bao {\it et al.} \cite{00Bao} derived an estimate of 482$\pm$92 mb. In contrast, adopting  $\langle \sigma^{Maxw}_{\gamma} (30 keV) \rangle$ = 266$\pm$5 mb for $^{93}$Nb \cite{00Bao}, Mughabghab calculated  $\langle \sigma^{Maxw}_{\gamma} (30 keV) \rangle$ = 314$\pm$63 mb by the detailed method, and 295$\pm$59 mb by the comparative one. The average of these results, {\it i.e.} 305$\pm$43 mb, agrees with JENDL-3.3 and the other data libraries.
\begin{figure}
\begin{center}
\includegraphics[height=6cm]{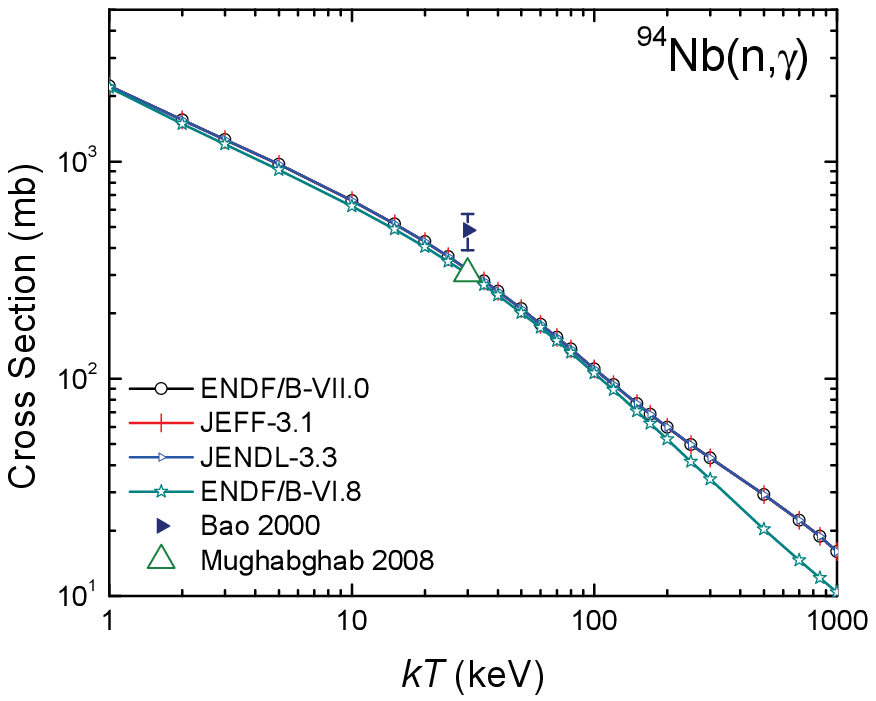}
\caption{Evaluated MACS for $^{94}$Nb \cite{00Bao,08Mugh}.}
\label{94Nb}
\end{center}
\end{figure}

{\bf $^{95}$Nb} (T$_{1/2}$ = 34.99 d): There are no experimental data, nor average resonance parameter information for $^{95}$Nb. The evaluated data libraries give 404 mb for  $\langle \sigma^{Maxw}_{\gamma} (30 keV) \rangle$ while Bao {\it et al.'s} systematics estimate is 310$\pm$65 mb \cite{00Bao}. Relying on the comparative method, Mughabghab \cite{08Mugh} calculated a value of 346$\pm$80 mb for this quantity, a result consistent with the other values within the uncertainty limits.

{\bf $^{98}$Ru}: Bao {\it et al.} recommend  $\langle \sigma^{Maxw}_{\gamma} (30 keV) \rangle$=173$\pm$36 mb \cite{00Bao} derived through systematics. The corresponding values for the libraries, ENDF/B-VII.0, JENDL-3.3 and JEFF-3.3 are, respectively,  237 mb, 237 mb and 125 mb. Relative to the $^{100}$Ru  $\langle \sigma^{Maxw}_{\gamma} (30 keV) \rangle$=206$\pm$13 mb \cite{82Mac}, Mughabghab \cite{08Mugh} derived 258$\pm$54 mb which is consistent with the JENDL-3.3 and ENDF/B-VII.0 values calculated by the CASTHY code \cite{75Ig}, but not the JEFF-3.3 result. Within the quoted experimental uncertainties, this value is consistent with Bao {\it et al.'s} recommendation \cite{00Bao}.

{\bf $^{103}$Ru} (T$_{1/2}$ = 39.3 d): Resonance parameters up to 333 eV are known  for $^{103}$Ru  \cite{06Mugh}, while the average resonance parameters remain undetermined. Another important consideration for $^{103}$Ru is the fact that its inelastic-neutron threshold opens at 2.8 keV, well below 30 keV. Consequently, this open channel must be considered. From their systematics analysis, Bao {\it et al.} \cite{00Bao} computed  $\langle \sigma^{Maxw}_{\gamma} (30 keV) \rangle$=343$\pm$52 mb. The JENDL-3.3, JEFF-3.1, and ENDF/B-VII.0 evaluations produce their corresponding values of 580 mb, 1510 mb , and 580 mb. The average resonance parameters for $^{103}$Ru are deduced relative to those of $^{101}$Ru \cite{08Mugh}. Using systematics, and adopting a 30-keV value of 996 mb for $^{101}$Ru, Mughabghab \cite{08Mugh} calculates   $\langle \sigma^{Maxw}_{\gamma} (30 keV) \rangle$=534 mb for $^{103}$Ru. However, detailed analysis revealed a value of 646 mb \cite{08Mugh}. The average of these results is 590$\pm$56 mb. Our present findings show that the previous recommendation, 343$\pm$52 mb \cite{00Bao} is a high underestimate.

{\bf $^{115m}$Cd} (T$_{1/2}$ = 44.56 d): There are no experimental cross-section data for $^{115m}$Cd;  ENDF/B-VII.0 gives $\langle \sigma^{Maxw}_{\gamma} (30 keV) \rangle$=225 mb, while Bao {\it et al.} \cite{00Bao} recommend 601$\pm$200 mb, based on systematics. To resolve this uncertainty, Mughabghab \cite{08Mugh} first followed the comparative method and determined this capture cross section as $\langle \sigma^{Maxw}_{\gamma} (30 keV) \rangle$=385 mb relative to the $^{111}$Cd value of 754$\pm$12 mb \cite{06Mugh}. Since resonance parameters are not known for isotope $^{115m}$Cd, its average resonance parameters are deduced from those of $^{111}$Cd based on systematics \cite{06Mugh,08bMugh}. Aided by these quantities, the detailed method gave a corresponding value of 361 mb. Together, they afford an average value of 373$\pm$80 mb; the uncertainty largely reflects that of the average level spacings.

{\bf $^{106,108,116}$Cd}: Evaluated nuclear-libraries values for $^{106,108,116}$Cd are based on experimental results, such as the data of Musgrove {\it et al.} \cite{78De} while Bao {\it et al.} \cite{00Bao} used the data from Theis {\it et al.} \cite{98Th}. $^{116}$Cd (T$_{1/2}$ = 3.0$\times$10$^{19}$ y), evaluated via the nuclear library is higher than Bao {\it et al.'s} number \cite{00Bao} and  consistent with that in {\it Atlas} \cite{06Mugh} and the recent measurement of Wisshak {\it et al.} \cite{02Wi}.  

\begin{figure}
\begin{center}
\includegraphics[height=6cm]{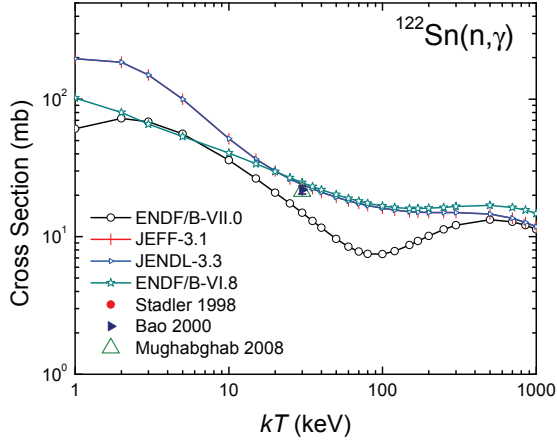}
\caption{Evaluated MACS and experimental cross sections for $^{122}$Sn \cite{00Bao,08Mugh,98Sta}.}
\label{122Sn}
\end{center}
\end{figure}

{\bf $^{122}$Sn}: Bao {\it et al.'s} recommended value \cite{00Bao}, 21.9$\pm$1.5 mb, comes from an unpublished thesis by Stadler \cite{98Sta}. The ENDF/B-VII.0-, JEFF-3.1-, and JENDL-3.3, respectively, give 14.9 mb, 23.9 mb, and 23.9 mb. The first library adopted the {\it Atlas} parameters \cite{06Mugh} that extended to 300 keV for the resolved-resonance region. In this library, the average $s$- and $p$-wave radiative widths were assumed to be 34 meV and 77 meV, respectively, founded on systematics from neighboring nuclei. In a recent detailed analysis, Mughabghab derived the corresponding values of 44$\pm$4 meV, and 83.8$\pm$3.8 meV, along with a $d$-wave capture width of 58$\pm$4 meV \cite{08Mugh}. To avoid missing the contribution of weak $p$-wave resonances, we specified the unresolved-energy region considered in this analysis as 10 keV - 2 MeV. With these parameters, the  $\langle \sigma^{Maxw}_{\gamma} (30 keV) \rangle$ is calculated as 21.3$\pm$2.0 mb, in excellent agreement with Stadler's result \cite{98Sta}. The JENDL-3.3 value, 23.9 mb, is slightly overestimated since the CASTHY calculations \cite{75Ig} on which it was based adopted an average $s$- and  $p$-wave radiative widths of 130 meV, as shown in Fig. \ref{122Sn}.

{\bf $^{125}$Sn} (T$_{1/2}$ = 9.64 d): There are no resonance parameter information and cross section measurements for this nucleus. Hence, the determination of the capture cross section relies on systematics. Bao {\it et al.} estimate a  $\langle \sigma^{Maxw}_{\gamma} (30 keV) \rangle$=59$\pm$9 mb \cite{00Bao}. However, the ENDF/B-VII.0 evaluation yields 98.3 mb. Applying the comparative method Mughabghab \cite{08Mugh}  obtained 58 mb relative to $^{117}$Sn $\langle \sigma^{Maxw}_{\gamma} (30 keV) \rangle$=318.8 mb \cite{86Wis},  while the detailed method computed a value of 78 mb. We recommend using a value of 68$\pm$20 mb that reflects additional uncertainties in the input parameters, particularly the derived average-level spacing of the $^{125}$Sn resonances.  

{\bf $^{125}$Sb} (T$_{1/2}$ = 2.76 y): In the absence of neutronic information for this isotope, systematics allow us to estimate its neutron-capture cross section. The ENDF/B-VII.0 evaluation, borrowed from JENDL-3.3, generates 527 mb for $\langle \sigma^{Maxw}_{\gamma} (30 keV) \rangle$, while ENDF/B-VI.8 gives 322 mb. Bao {\it et al.'s} \cite{00Bao} estimate is 260$\pm$70 mb. Applying systematics, Mughabghab \cite{08Mugh} calculated 190 mb by the comparative method adopting $\langle \sigma^{Maxw}_{\gamma} (30 keV) \rangle$=303$\pm$9 mb for $^{123}$Sb \cite{93Kae}, and 252 mb by the detailed cross section method, so resulting in an average value of 221$\pm$23 mb. This result shows very good agreement within the uncertainty limits with  Bao {\it et al.'s} \cite{00Bao} estimate, but departs substantially from the ENDF/B-VII.0 and JENDL-3.3 value of 527 mb. 

{\bf $^{120}$Te}: Information on the resonance parameters is not available for this isotope with a natural element abundance of 0.091\%. Consequently, we resort to systematics to appraise its capture cross section. The $\langle \sigma^{Maxw}_{\gamma} (30 keV) \rangle$ values from the libraries of ENDF/B-VII.0, JEFF-3.1, and JENDL-3.3 are 292 mb, 430 mb, and 292 mb, respectively. In contrast, Bao {\it et al.} reported a value of 420$\pm$103 mb \cite{00Bao}. Applying both the detailed- and systematics-methods, Mughabghab \cite{08Mugh} obtained 428$\pm$50 mb, a value in agreement with that of the JEFF-3.1 library, which was adopted from the ENDF/B-VI.8 library, Bao {\it et al.'s} estimate \cite{00Bao}, and Dillman's recent measurement of this isotope \cite{06Dill}. In the systematic method, Mughabghab \cite{08bMugh} derived the average resonance parameters of $^{120}$Te relative to those of $^{122}$Te, and adopted $\langle \sigma^{Maxw}_{\gamma} (30 keV) \rangle$ = 280$\pm$10 mb for $^{122}$Te \cite{06Mugh}.

{\bf $^{129}$Xe}: The 30-keV MACS values computed from the data libraries and the Bao {\it et al.} estimate \cite{00Bao} range from 421 mb to 478 mb. However, Reifarth {\it et al.} \cite{02Re} reported a measurement of 617$\pm$12 mb for this quantity, which was adopted in the {\it Atlas} \cite{06Mugh}. The cause of the variance is that the measured capture cross section  \cite{02Re} is high compared with the evaluated libraries. In particular, this factor is 46\% at 30 keV.

{\bf $^{130}$Xe}: The ENDF/B-VII.0-, JEFF-3.1-, and JENDL-3.3-derived 30-keV MACS are 153 mb, 152 mb, and 283 mb, respectively. Two measurements for this quantity are reported in the literature: one by the prompt gamma method (132$\pm$2.1 mb) \cite{02Re}, the other by the activation method (192$\pm$54 mb) \cite{91Be}; the {\it Atlas} value \cite{06Mugh} is based on these two. From the resolved- and unresolved-resonance parameters, and guided by the capture cross section of Reifarth {\it et al.} \cite{02Re}, Mughabghab \cite{08Mugh} computes for this nucleus  $\langle \sigma^{Maxw}_{\gamma} (30 keV) \rangle$=137$\pm$6 mb.  

\begin{figure}
\begin{center}
\includegraphics[height=6cm]{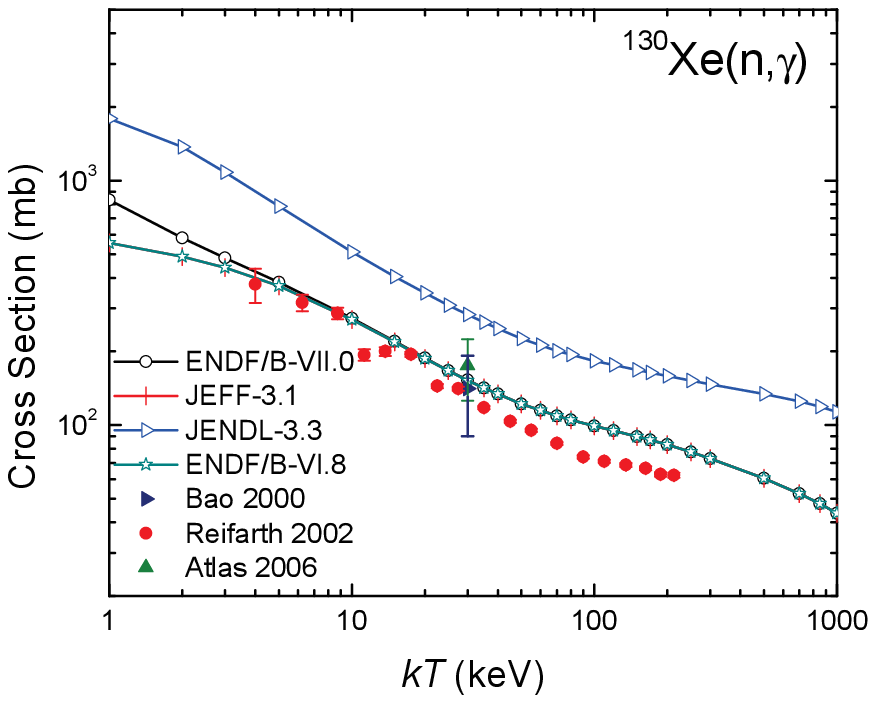}
\caption{Evaluated MACS and experimental cross sections for $^{130}$Xe \cite{00Bao,06Mugh,02Re}.}
\label{130Xe}
\end{center}
\end{figure}

Fig. \ref{130Xe} graphically shows that $^{130}$Xe evaluation in JENDL-3.3 library clearly overestimates experimental $\sigma (n,\gamma)$, while the values in the ENDF/B-VII.0 and JEFF-3.1 libraries are consistent with the experiment. This fact, consistent with the  previously observed deviation between the ENDF/B-VII.0 and JENDL-3.3 libraries for  the product of  $\langle \sigma^{Maxw}_{\gamma} (30 keV) \rangle$ and solar-system abundances, explains the disagreement between the findings of Bao {\it et al.} \cite{00Bao} and Nakagawa {\it et al.} \cite{05Nak}.

{\bf $^{136}$Xe}: For the $N$=82 nucleus $^{136}$Xe, JENDL-3.3 evaluation agrees well with the Bao {\it et al.} value \cite{00Bao} while {\it Atlas} \cite{06Mugh} provides somewhat lower numbers than ENDF/B-VII.0 library and the measurements of Beer \cite{91Be} and Leist  {\it et al.} \cite{85Le}.

{\bf $^{134}$Cs} (T$_{1/2}$ = 2.07 y): There is a large discrepancy between the cross section values derived from ENDF/B-VII.0 (JENDL-3.3), 1160 mb, and JEFF-3.1, 588 mb, that can be attributed to the use of different computer codes and input parameters. The estimate made by Bao {\it et al.}, 664$\pm$174 mb \cite{00Bao}, conforms with that of  JEFF-3.1. To resolve this discrepancy, Mughabghab \cite{08Mugh} used the resolved parameters of $^{134}$Cs \cite{06Mugh}, and the unresolved parameters derived from the systematics for his assessment \cite{06Mugh,08bMugh}. The value he obtained was 530$\pm$160 mb, agreeing with Bao {\it et al.'s} estimate \cite{00Bao}, the recent measurements of Patronis {\it et al.} \cite{04Pa}, and the JEFF-3.1 evaluation but disagreeing with the ENDF/B-VII.0 (JENDL-3.3) evaluation in the unresolved-energy region.

\begin{figure}
\begin{center}
\includegraphics[height=6cm]{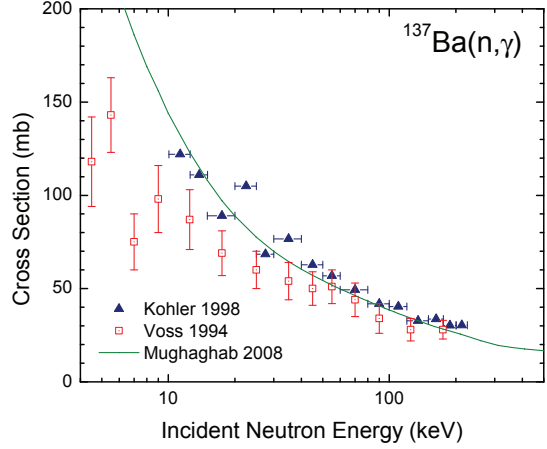}
\caption{Calculated and experimental cross sections for $^{137}$Ba \cite{98Ko,94Vo}. The solid curve is a theoretical calculation of the capture cross section in terms of the Lane and Lynn formalism \cite{57Lan}, with the following neutron resonance parameters: $s$- and $p$-wave neutron (gamma) strength functions (in units of 10$^{-4}$) 0.78(4.07), and 1.5 (5.37), respectively \cite{08Mugh}.}
\label{137Ba}
\end{center}
\end{figure}

{\bf $^{137}$Ba}: It is no surprise that  $\langle \sigma^{Maxw}_{\gamma} (30 keV) \rangle$ derived from the evaluated data libraries give a value close to that reported by Musgrove {\it et al.}, 58$\pm$10 mb \cite{76Mu}, since the $\sigma (n,\gamma)$ of these libraries in the unresolved-energy region were fitted to the Musgrove measurement.The value recommended by Bao {\it et al.}, 76.3$\pm$2.4 mb \cite{00Bao}, is an average of the reported values by Koehler {\it et al.} \cite{98Ko} and Voss {\it et al.} \cite{94Vo}. To provide additional clarity in this case, Mughabghab \cite{08Mugh} calculated the $\sigma (n,\gamma)$ in the unresolved-resonance region using the systematics of the average resonance parameters \cite{06Mugh,08bMugh}. From these results, depicted in Fig. \ref{137Ba}, we conclude
\begin{description}
\item [i] There is good agreement between Mughabghab's calculation and the Koehler {\it et al.'s} ORNL measurement \cite{98Ko}.
\item [ii] Voss {\it et al.'s}  measurement (KFK) \cite{94Vo} is problematic at the energy region below 12 keV \cite{98Ko}.  
\end{description}
From our current calculation, the  $\langle \sigma^{Maxw}_{\gamma} (30 keV) \rangle$=72.3$\pm$2.4 mb for $^{137}$Ba.

{\bf $^{142}$Nd}: The ENDF/B-VII.0 library agrees well with Bao {\it et al.} \cite{00Bao}, {\it Atlas} \cite{06Mugh} and Wisshak {\it et al.'s} data \cite{98Wi} for $^{142}$Nd, while the JENDL-3.3 library has a much higher $\sigma (n,\gamma)$ value. This large JENDL-3.3 cross section value for $^{142}$Nd evaluation is consistent with the previously observed problem for the product of  $\langle \sigma^{Maxw}_{\gamma} (30 keV) \rangle$ and solar-system abundances. 
\begin{figure}
\begin{center}
\includegraphics[height=6cm]{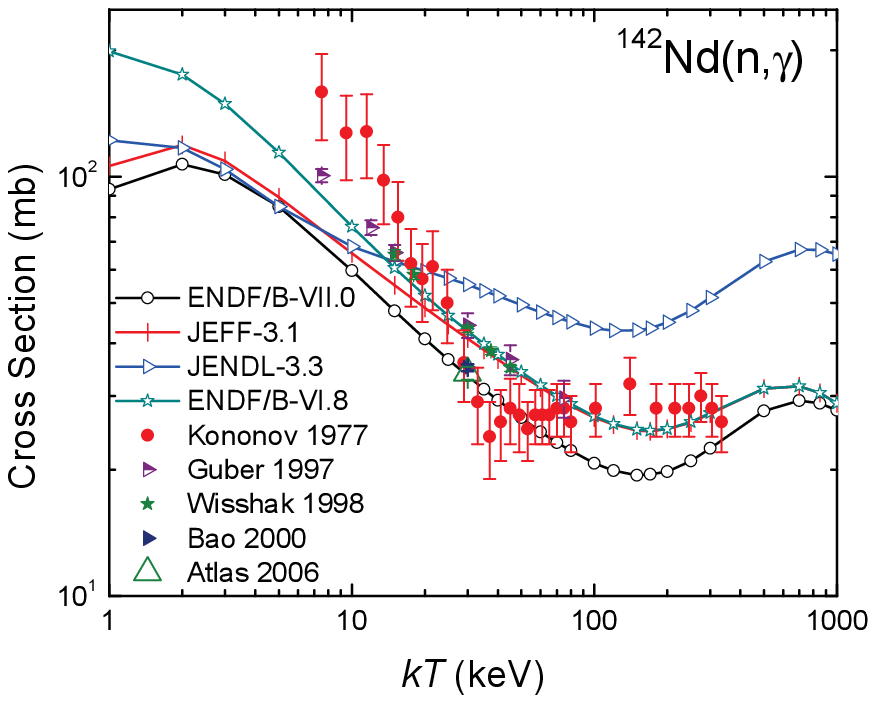}
\caption{Evaluated MACS and experimental cross sections for $^{142}$Nd \cite{00Bao,06Mugh,77Ko,97Gu,98Wi}.}
\label{142Nd}
\end{center}
\end{figure}

{\bf $^{148g,149}$Pm}: The ENDF/B-VII.0 and JENDL-3.3 libraries produce lower cross section values than systematic approach of Bao {\it et al.} \cite{00Bao} for $^{148g}$Pm (T$_{1/2}$ = 5.4 d) and $^{149}$Pm ( T$_{1/2}$ = 53.1 h) that agrees well with the recent measurement of Reifarth {\it et al.} \cite{03Rei}. 

{\bf $^{159}$Tb}: Evaluated nuclear libraries, with the possible exception of ENDF/B-VI.8, overestimate the cross section values for  $^{159}$Tb. Lower cross section values are consistent with the results of Bao {\it et al.} \cite{00Bao}, {\it Atlas} \cite{06Mugh} and the measurements of Bokhovko {\it et al.} \cite{96Bo}.

{\bf $^{162,164}$Er}: Experimental data are limited for these isotopes \cite{89Tr,01Be} and do not allow a definite conclusion for $^{162}$Er. The $^{164}$Er(n,$\gamma$) cross section measurement of Best {\it et al.} {\cite{01Be,96Bes}  adopted by Bao {\it et al.} \cite{00Bao}, {\it Atlas} \cite{06Mugh}, and Mughabghab's calculations \cite{08Mugh} provide lower values than those of the evaluated libraries.

{\bf $^{208}$Pb}: Up to two thirds of the MACS in the doubly magic nucleus $^{208}$Pb comes from direct capture. The {\it Atlas} data \cite{06Mugh} clearly recommends Bao {\it et al.'s} \cite{00Bao}  $\langle \sigma^{Maxw}_{\gamma} (30 keV) \rangle$ for this nucleus; this value is in concert with the experimental data \cite{97Be,03Be,04Ra} and provides lower than the nuclear libraries cross section values shown in Fig. \ref{208Pb}.
\begin{figure}
\begin{center}
\includegraphics[height=6cm]{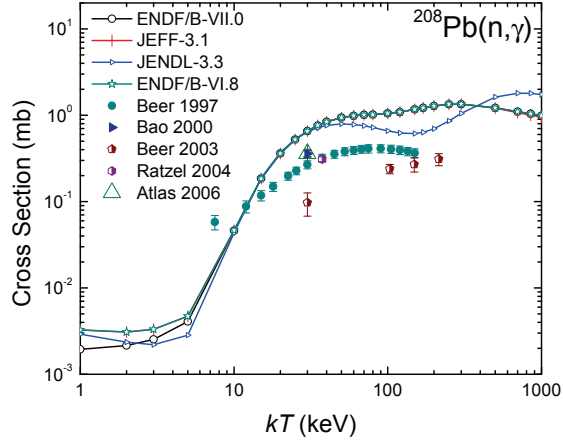}
\caption{Evaluated MACS and experimental cross sections for $^{208}$Pb \cite{00Bao,06Mugh,97Be,03Be,04Ra}.}
\label{208Pb}
\end{center}
\end{figure}

Table \ref{Bao} summaries the results and recommendations for   $\langle \sigma^{Maxw}_{\gamma} (30 keV) \rangle$. They indicate reasonably good agreement between the ENDF evaluations, {\it Atlas of Neutron Resonances} \cite{06Mugh}, and the MACS recommendations of Bao {\it et al.} \cite{00Bao} and KADONIS \cite{06Dil}. Several discrepancies between  ENDF libraries, the calculations of Mughabghab \cite{08Mugh} and recommendations of Bao {\it et al.} \cite{00Bao} indicate the need for improving nuclear-data sets, and  undertaking new measurements, marked as $\dagger$ and $^{\star}$, respectively.   

\subsubsection{(n,fission) Reactions}
The ENDF libraries are the primary source of neutron-fission data for nuclear-engineering applications. Table \ref{MACSTable18} presents the MACS and the reaction rates for neutron-fission reactions from the four major evaluated libraries. An extensive analysis of experimental- and evaluated-fission data recommends the JENDL-3.3 library MACS for  $^{240-243,248}$Cm, $^{247}$Bk, $^{250}$Cf, and the ENDF/B-VII.0 library  cross sections for the rest of the fission nuclei.

\subsubsection{(n,p), (n,$\alpha$) and (n,2n) Reactions}
(n,p), (n,$\alpha$) and (n,2n) reactions have been calculated and the results are available at the NNDC Web server {\it (http://www.nndc.bnl.gov/astro)} \cite{06Pri}. A limited amount of experimental data, in the 1 keV - 1 MeV region, restricts comprehensive data-analysis. However, we observed data deficiencies in evaluating $^{22}$Na and $^{16}$O, $^{123}$Te for (n,p) and (n,$\alpha$) reactions, respectively. Comparing the experimental- and evaluated-data sets for these cases indicates problems in underestimating cross sections in the  evaluated neutron libraries.

\subsection{Astrophysical Reaction Rates}

Astrophysical reaction rates were calculated in parallel with the MACS using Eq. (\ref{myeq.a6}).  Fig. \ref{fig6} gives the ENDF/B-VII.0 reaction rates at {\it kT}=30 keV. The values are function of the MACS and mean thermal velocities. Therefore, the basic shape of the reaction rates as a function of the ENDF material (nucleus) is very similar to that of the MACS plotted in Figure \ref{fig5}. 
\begin{figure}[ht!]
\begin{center}
\includegraphics[height=7cm]{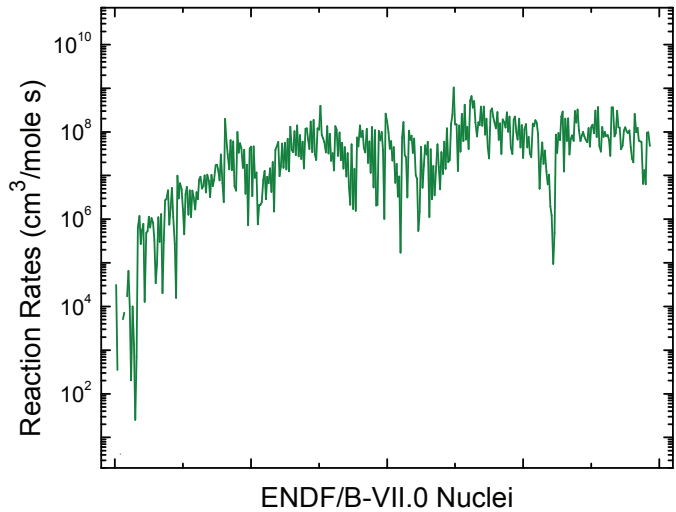}
\caption{Astrophysical reaction rates for ENDF/B-VII.0 evaluated neutron library at {\it kT=30} keV.}
\label{fig6}
\end{center}
\end{figure}

\section{Conclusions}

We completed a  large-scale calculation of Maxwellian-averaged cross sections  and astrophysical reaction rates  using the ENDF/B-VII.0-, JEFF-3.1-, JENDL-3.3-, and ENDF/B-VI.8-evaluated nuclear reaction libraries from 1 keV to 1 MeV.  The calculated cross sections and reaction rates were compared with the existing JENDL-3.3 calculations, the published recommendations \cite{00Bao,00Rau,06Mugh,05Nak,06Dil}, the recent  calculations of Mughabghab \cite{08Mugh} and experimental data \cite{exfor}. In most cases, the data sets are in agreement, but discrepancies were observed. A combined analysis of several data-sources indicates that Nakagawa {\it et al.'s} findings \cite{05Nak} are well-reproducible, and an explanation for the discrepancies between the JENDL-3.3 cross section values and the recommendations of Bao {\it et al.} \cite{00Bao} lies in the quality of nuclear-reaction cross section data.

Our work is a part of ongoing validation of ENDF/B-VII.0, and provides complementary information on deficiencies in evaluated nuclear-reaction data  in the keV energy region. This work will lead to the reevaluation of the obsolete data sets in the next ENDF/B-VII.1 library release, along with additional measurements and better contacts between the ENDF- and nuclear astrophysics-communities. Further improvements in the ENDF libraries and the development of nuclear-data covariances  \cite{08Sa,08Li,08Her} will entail the better agreement between two approaches that will benefit both nuclear astrophysics and nuclear-industry applications.

\ack
The authors thanks Drs. F. Kaeppeler and I. Dillmann for productive discussions and helpful suggestions.  We also are grateful to Drs. M.W. Herman, and A.D. Woodhead and Mrs. M.T. Blennau for  the help with ENDF utility codes, and carefully reading the manuscript, respectively. This work was funded by the Office of Nuclear Physics, Office of Science of the U.S. Department of Energy, under Contract No. DE-AC02-98CH10886 with Brookhaven Science Associates, LC.  

\begin{appendix}

\section{Supplemental data}
Supplementary data sets associated with this article can be found in the on-line version, at doi:X.X/j.adt.X.X.X.

\end{appendix}



\newpage

\section*{EXPLANATION OF TABLES}\label{sec.eot}
\addcontentsline{toc}{section}{EXPLANATION OF TABLES}

{\bf Table I.} 
Maxwellian-averaged (n,$\gamma$) cross sections (mbarns) at $kT$=30 keV from the evaluated nuclear-reaction libraries, Atlas of Neutron Resonances \cite{06Mugh}, Mughabghab's calculation \cite{08Mugh}, Bao {\it et al.'s} and KADONIS's recommended values \cite{00Bao,06Dil}.

ENDF/B-VII.0: Evaluated Nuclear Data File, U.S. 2006 \cite{06Chad}

JEFF-3.1: Joint European Fission and Fusion File, Europe 2005 \cite{02Jac,04Kon}

JENDL-3.3: Japanese Evaluated Nuclear Data Library, Japan 2002 \cite{02Shi}

ENDF/B-VI.8: Evaluated Nuclear Data File, U.S. 2001 \cite{01Cse}




{\bf Table II.} Maxwellian-averaged (n,$\gamma$) cross sections and reaction rates from the evaluated nuclear reaction libraries.

{\bf Table III.} Maxwellian-averaged (n,fission) cross sections and reaction rates from the evaluated nuclear reaction libraries.

\newpage
\datatables
\setlength{\LTleft}{0pt}
\setlength{\LTright}{0pt} 


\setlength{\tabcolsep}{0.5\tabcolsep}

\renewcommand{\arraystretch}{1.0}
{\footnotesize
\begin{longtable}{|l|l|l|l|l|l|l|l|l|}
\caption[ Maxwellian-averaged (n,$\gamma$) cross sections (mbarns) at $kT$=30 keV from the evaluated nuclear-reaction libraries, Atlas of Neutron Resonances, Mughabghab's calculation, Bao {\it et al.'s} and KADONIS's  recommended values.]{ Maxwellian-averaged (n,$\gamma$) cross sections (mbarns) at $kT$=30 keV from the evaluated nuclear-reaction libraries, Atlas of Neutron Resonances \cite{06Mugh}, Mughabghab's calculation \cite{08Mugh}, Bao {\it et al.'s} and KADONIS's recommended values \cite{00Bao,06Dil}.} \label{Bao} \\
\hline 
\multicolumn{1}{|c|}{\textbf{Nucleus}} & 
\multicolumn{1}{c|}{\textbf{ENDF/B-VII.0}} & 
\multicolumn{1}{c|}{\textbf{JEFF-3.1}}& 
\multicolumn{1}{c|}{\textbf{JENDL-3.3}}& 
\multicolumn{1}{c|}{\textbf{ENDF/B-VI.8}}& 
\multicolumn{1}{c|}{\textbf{Atlas of Neutron}}& 
\multicolumn{1}{c|}{\textbf{Mughabghab}}& 
\multicolumn{1}{c|}{\textbf{Bao {\it et al.}}}& 
\multicolumn{1}{c|}{\textbf{KADONIS}} \\ 
\multicolumn{1}{|c|}{\textbf{}} & 
\multicolumn{1}{c|}{\textbf{}} & 
\multicolumn{1}{c|}{\textbf{}}& 
\multicolumn{1}{c|}{\textbf{}}& 
\multicolumn{1}{c|}{\textbf{}}& 
\multicolumn{1}{c|}{\textbf{Resonances }}& 
\multicolumn{1}{c|}{\textbf{}}& 
\multicolumn{1}{c|}{\textbf{}}& 
\multicolumn{1}{c|}{\textbf{}} \\
\hline \endfirsthead \multicolumn{9}{l}{{\bfseries \tablename\ \thetable{}} --  Maxwellian-averaged (n,$\gamma$) cross sections (mbarns) at $kT$=30 keV.} \\ 
\hline 
\multicolumn{1}{|c|}{\textbf{Nucleus}} & 
\multicolumn{1}{c|}{\textbf{ENDF/B-VII.0}} & 
\multicolumn{1}{c|}{\textbf{JEFF-3.1}}& 
\multicolumn{1}{c|}{\textbf{JENDL-3.3}}& 
\multicolumn{1}{c|}{\textbf{ENDF/B-VI.8}}& 
\multicolumn{1}{c|}{\textbf{Atlas of Neutron}}& 
\multicolumn{1}{c|}{\textbf{Mughabghab}}& 
\multicolumn{1}{c|}{\textbf{Bao {\it et al.} }} & 
\multicolumn{1}{c|}{\textbf{KADONIS}}\\ 
\multicolumn{1}{|c|}{\textbf{}} & 
\multicolumn{1}{c|}{\textbf{}} & 
\multicolumn{1}{c|}{\textbf{}}& 
\multicolumn{1}{c|}{\textbf{}}& 
\multicolumn{1}{c|}{\textbf{}}& 
\multicolumn{1}{c|}{\textbf{Resonances}}& 
\multicolumn{1}{c|}{\textbf{}}& 
\multicolumn{1}{c|}{\textbf{}} & 
\multicolumn{1}{c|}{\textbf{}}\\
\hline 
\endhead \hline \multicolumn{9}{|r|}{{Continued on next page}} \\ 
\hline \endfoot  \hline  \endlastfoot
1-H-1 & 1.53E-1 & 1.53E-1 & 1.53E-1 & 1.53E-1 &  & & 0.254$\pm$0.020 & 0.254$\pm$0.020 \\
1-H-2 & 1.99E-3 & 1.99E-3 & 2.35E-3 & 1.99E-3 &  & &  & \\
1-H-3 &  &  &  &  &  & &  &  \\
2-He-3 & 2.48E-5$\dagger$ & 2.48E-5$\dagger$ & 1.63E-2 & 2.48E-5$\dagger$ & 0.0077$\pm$0.0007 & & 0.0076$\pm$0.0006 & 0.0076$\pm$0.0006  \\
2-He-4 &  &  &  &  &  & &  &  \\
3-Li-6 & 3.29E-2 & 3.29E-2 & 3.35E-2 & 3.29E-2 &  & &   & \\
3-Li-7 & 4.67E-2 & 4.67E-2 & 3.92E-2 & 4.67E-2 & 0.0393$\pm$0.003 & & 0.042$\pm$0.003 & 0.042$\pm$0.003 \\
4-Be-9 & 1.13E-1 & 1.27E-2 & 6.66E-3 & 1.13E-2 &  & &  & \\
\hline
\end{longtable}
}
\begin{description}
\item [$\star$]  New measurements are necessary to resolve the discrepancy between evaluated nuclear libraries, calculation of Mughabghab \cite{08Mugh} and Bao {\it et al.} recommendation \cite{00Bao}.
\item [$\dagger$] (n,$\gamma$) section of the evaluated nuclear reaction library needs an update.
\item [*] From systematics \cite{08Mugh}.
\item [**] NON-SMOKER calculations re-normalized to nearby experimental data \cite{00Bao}.
\item [***] Theoretical data in KADONIS.
\item [a] Atlas bound level \cite{06Mugh}.
\item [b] BNL-325 bound level \cite{81BNL325}.
\item [T] Misprint in the Atlas.
\end{description}

{\footnotesize 
\newpage
\begin{longtable}{|l|c|l|l|l|l|l|l|l|l|l|l|}
\caption[Maxwellian-averaged (n,$\gamma$) cross sections and astrophysical reaction rates]{Maxwellian-averaged (n,$\gamma$) cross sections and astrophysical reaction rates.} \label{MACSTable102} \\
\endfirsthead
\multicolumn{12}{l}%
{{\bfseries \tablename\ \thetable{}} -- Maxwellian-averaged (n,$\gamma$) cross sections and astrophysical reaction rates.} \\
\endhead
\multicolumn{12}{|r|}{{Continued on next page}} \\ \hline
\endfoot
\endlastfoot
\hline  \multicolumn{4}{|r|}{1-H-1}   &  \multicolumn{2}{|r|}{1-H-2}   & \multicolumn{2}{|r|}{2-He-3} & \multicolumn{2}{|r|}{3-Li-6} & \multicolumn{2}{|r|}{3-Li-7} \\ \hline 
 {\it kT} (keV)& Library &  Sigma (mb) & R ($T_{9}$)  &  Sigma (mb) & R ($T_{9}$)  &  Sigma (mb) & R ($T_{9}$)  &  Sigma (mb) & R ($T_{9}$)   &  Sigma (mb) & R ($T_{9}$) \\ \hline 
1 &  ENDF/B-VII.0  & 1.163E0 & 4.335E4  & 2.094E-3 & 6.758E1  & 1.358E-4 & 4.133E0 & 1.804E-1 & 5.134E3 & 2.112E-1 & 5.95E3 \\
 &  JEFF-3.1  & 1.182E0 & 4.403E4  & 2.094E-3 & 6.758E1  & 1.358E-4 & 4.133E0 & 1.804E-1 & 5.134E3 & 2.112E-1 & 5.95E3 \\
 &  JENDL-3.3  & 1.182E0 & 4.403E4  & 2.634E-3 & 8.498E1  & 3.619E-3 & 1.101E2 & 1.802E-1 & 5.13E3 & 2.149E-1 & 6.053E3 \\
 &  ENDF/B-VI.8  & 1.182E0 & 4.403E4  & 2.094E-3 & 6.758E1  & 1.358E-4 & 4.133E0 & 1.804E-1 & 5.134E3 & 2.112E-1 & 5.95E3 \\
5 &  ENDF/B-VII.0  & 4.879E-1 & 4.065E4  & 1.296E-3 & 9.351E1  & 6.075E-5 & 4.133E0 & 8.064E-2 & 5.133E3 & 1.161E-1 & 7.313E3 \\
 &  JEFF-3.1  & 4.851E-1 & 4.042E4  & 1.296E-3 & 9.351E1  & 6.075E-5 & 4.133E0 & 8.064E-2 & 5.133E3 & 1.161E-1 & 7.313E3 \\
\end{longtable}
}

{\footnotesize 
\newpage
\begin{longtable}{|l|c|l|l|l|l|l|l|l|l|l|l|}
\caption[Maxwellian-averaged (n,fission) cross sections and astrophysical reaction rates]{Maxwellian-averaged (n,fission) cross sections and astrophysical reaction rates.} \label{MACSTable18} \\
\endfirsthead
\multicolumn{12}{l}%
{{\bfseries \tablename\ \thetable{}} -- Maxwellian-averaged (n,fission) cross sections and astrophysical reaction rates} \\
\endhead
\multicolumn{12}{|r|}{{Continued on next page}} \\ \hline
\endfoot
\endlastfoot
\hline  \multicolumn{4}{|r|}{88-Ra-223}   &  \multicolumn{2}{|r|}{88-Ra-226}   & \multicolumn{2}{|r|}{89-Ac-227} & \multicolumn{2}{|r|}{90-Th-227} & \multicolumn{2}{|r|}{90-Th-228} \\ \hline 
 {\it kT} (keV)& Library &  Sigma (mb) & R ($T_{9}$)  &  Sigma (mb) & R ($T_{9}$)  &  Sigma (mb) & R ($T_{9}$)  &  Sigma (mb) & R ($T_{9}$)   &  Sigma (mb) & R ($T_{9}$) \\ \hline 
1 &  ENDF/B-VII.0  & 6.077E1 & 1.604E6  & 9.906E-5 & 2.615E0  & 2.46E-3 & 6.495E1 & 2.489E3 & 6.572E7 & 1.252E0 & 3.305E4 \\
 &  JEFF-3.1  & 6.077E1 & 1.604E6  & 9.906E-5 & 2.615E0  & 2.46E-3 & 6.495E1 & 2.489E3 & 6.572E7 & 1.252E0 & 3.305E4 \\
 &  JENDL-3.3  & 6.077E1 & 1.604E6  & 9.906E-5 & 2.615E0  & 2.46E-3 & 6.495E1 & 2.489E3 & 6.572E7 & 1.252E0 & 3.305E4 \\
 & ENDF/B-VI.8 &  &  &  &  &  &  &  &  &  &  \\
5 &  ENDF/B-VII.0  & 6.166E1 & 3.64E6  & 3.902E-4 & 2.303E1  & 8.567E-3 & 5.057E2 & 1.451E3 & 8.568E7 & 1.362E-1 & 8.037E3 \\
 &  JEFF-3.1  & 6.166E1 & 3.64E6  & 3.902E-4 & 2.303E1  & 8.567E-3 & 5.057E2 & 1.451E3 & 8.568E7 & 1.362E-1 & 8.037E3 \\
\end{longtable}
}

\end{document}